\documentclass[aip, jmp,amsmath]{revtex4}

\usepackage{amssymb}
\usepackage[demo]{graphicx}
\usepackage{dcolumn}
\usepackage{bm}
\usepackage{amsmath}
\usepackage[utf8]{inputenc}
\usepackage[T1]{fontenc}
\usepackage{etoolbox}
\usepackage{enumerate}
\usepackage{color}
\usepackage{babel}
\usepackage{epstopdf}
\usepackage{float}
\usepackage{thmtools}
\usepackage{thm-restate}
\usepackage{hyperref}
\usepackage{cleveref}
\usepackage{tabularx}
\usepackage{tikz,subfig}
\usepackage{subfig}
\usepackage{caption}
\usepackage{times,euler,eucal,eufrak}

\usepackage{amsthm}

\DeclareMathOperator{\Tr}{Tr}

\DeclareMathOperator{\Wg}{Wg}
\DeclareMathOperator{\Id}{Id}

\newcommand{\bfG}{{\mathbf{G}}}
\newcommand{\bfX}{{\mathbf{X}}}
\newcommand{\bfV}{{\mathbf{V}}}
\newcommand{\bfE}{{\mathbf{E}}}
\newcommand{\bfu}{{\mathbf{u}}}
\newcommand{\bfv}{{\mathbf{v}}}
\newcommand{\bfs}{{\mathbf{s}}}
\newcommand{\bft}{{\mathbf{t}}}
\newcommand{\bfp}{{\mathbf{p}}}
\newcommand{\bfc}{{\mathbf{c}}}
\newcommand{\bff}{{\mathbf{f}}}
\newcommand{\bfx}{{\mathbf{x}}}

\newcommand{\alpt}{{\tilde{\alpha}}}
\newcommand{\bet}{{\tilde{\beta}}}

\newtheorem{thm}{Theorem}[section]
\newtheorem{cor}[thm]{Corollary}
\newtheorem{lem}[thm]{Lemma}
\newtheorem{prop}[thm]{Proposition}

\newtheorem{defn}{Definition}[section]
\newtheorem{rem}{Remark}[section]

\makeatletter
\def\@email#1#2{%
	\endgroup
	\patchcmd{\titleblock@produce}
	{\frontmatter@RRAPformat}
	{\frontmatter@RRAPformat{\produce@RRAP{*#1\href{mailto:#2}{#2}}}\frontmatter@RRAPformat}
	{}{}
}%
\makeatother
\begin{document}
	
	
	\title[Convergence of the quantum entropy of random graph states and its fluctuation]{Almost surely convergence of the quantum entropy of random graph states and the area law}
	\author{Z. Yin}
	\altaffiliation{School of Mathematics and Statistics, Central South University, China}%
	\author{L. Zhao}
	\altaffiliation{School of Mathematics, Harbin Institute of Technology, China}
	
	\date{\today}
	
	\begin{abstract}
		In \cite{BIK3}, Collins et al. showed that the quantum entropy of random graph states satisfies the so-called area law as the local dimension tends to be large.  
		In this paper, we continue to study the fluctuation of the convergence and thus prove the area law holds almost surely.  
	\end{abstract}

	\maketitle
	
	
	\section{Introduction}\label{Intro}

Entanglement is possibly one of the most intriguing phenomena among quantum systems. Describing the entanglement contained in a quantum system's state is generally complex when the state is highly entangled, as it requires a large number of parameters to describe it classically. In quantum many-body systems, analyzing the entanglement between two distinct subsystems, known as entanglement entropy or block entropy, is a meaningful avenue of research in various directions.
	
We consider a quantum lattice system. Let $\mathcal{G}$ be a lattice, and we associate an individual quantum system for each vertex $i$ in the lattice $\mathcal{G}.$ We divide $\mathcal{G}$ into two separable regions, denoted by $\mathcal{S}$ and $\mathcal{T},$ respectively. Suppose that the quantum lattice system is described by (local) quantum systems at the sites of region $\mathcal{S}.$  Let $|\psi\rangle$ be the pure ground state. Then the quantum state of the lattice system is given by the  reduced state $\rho_\mathcal{S} = \Tr_{\mathcal{T}}(|\psi\rangle\langle\psi|).$ We call the reduced state $\rho_\mathcal{S}$ satisfies an area law if its von Neumann entropy is proportional to the size of its boundary $\partial\mathcal{S},$ i.e.,
\begin{equation*}
H(\rho_{\mathcal{S}}) \propto |\partial\mathcal{S}|,
\end{equation*}
where $H(\rho_{\mathcal{S}})$ is the von Neumann entropy of $\rho_{\mathcal{S}}.$

The emergence of an area law requires that quantum correlations between the system and its exterior be established via its boundary surface \cite{ECP10}. For one-dimensional chains, Hastings proved the existence of an area law if the Hamiltonian of the system is gapped \cite{HMB07}. Later, Arad et al. provided a combinatorial proof for the special case of frustration-free systems \cite{ALV12}. Additionally, area laws are fulfilled if the quantum states of the systems satisfy the exponential decay of correlations \cite{BH13,BH15}. As well as the one-dimensional case, area laws have been found in high-dimensional systems \cite{PEDC05,CE06,CEPD06,ML09,NMJJ10,VKV13,FM15,MAVV16,GE16,AAG22,KI21}. Moreover, entanglement area laws for long-range interacting systems have also been considered \cite{GFBG17, KK20}. It was shown that an area law holds for any gapped local Hamiltonian over an arbitrary interaction (hyper) graph \cite{AHS22}. In \cite{TKS23}, experimental verification was provided that the equilibrium states of gapped quantum many-body systems satisfy the area law of quantum mutual information. However, there are quantum systems that do not fulfill the area law.

Hence, it is quite complicated to decide whether an area law holds for a quantum lattice system, and there is a vast amount of literature involved in this topic. Because describing the entanglement is quite complicated, and the parameters increase exponentially as the local dimension gets larger. Motivated by random matrix theory and its applications in quantum information theory \cite{BIK0, BI12, IN07, KK11, BF22, CYZ, MS17,ANV2016,G12,BI10,BI15,RIL21,CYZZ23,CGL23}, it is natural to sample the parameters of the ground state $|\psi\rangle$ randomly and to see whether an area law holds generically, since the randomness may provide more tools. 
In \cite{BIK3}, Collins et al. considered a rather general case: the quantum system is described by a graph (in the lattice), and they introduced the so-called random graph state; see Definition \ref{def-uni}. By using mathematical techniques in random matrix theory, they proved that the von Neumann entropy of the reduced state obeys an area law on average as the local dimension goes to be large. Namely, 
\begin{equation*}
\mathbb{E} [H(\rho_{\mathcal{S}})] \propto |\partial\mathcal{S}|,
\end{equation*}
where $\mathbb{E}$ is the expectation originated from the random structure. And they also conjecture that the above result holds true with high probability (or almost surely). 
In this paper, we will answer this question by improving their results from a probabilistic perspective; see our main result Theorem \ref{thm:almost-sure}. To this end, we will study the fluctuation of the empirical eigenvalue distribution of the reduced random graph state; see Proposition \ref{prop:almost surely}. 

We end this introduction with a brief description of the organization of the paper. In Section II, we present some preliminaries and notations. Section III is devoted to studying the covariance of the empirical eigenvalue distribution. By improving the estimations, we can prove the almost surely convergence of the empirical eigenvalue distribution, and thus the main result. Finally, in Section VI, we present some concrete examples. 
	


\section{Preliminaries}\label{Pre}

\subsection{Non-crossing partitions and permutations}
	
For any integer $n \geq 1,$ denote $[n]$ by the set $\{1,2,\cdots,n\}$. We say $\pi = \{V_1, V_2, \ldots, V_k\}$ is a partition of $[n]$ if the sets $V_i$ are disjoint and non-empty, and their union is equal to $[n].$ We call $V_1, \ldots, V_k$ the blocks of partition $\pi$. The set of all partitions of $[n]$ is denoted by $\mathcal{P}(n).$ 
	
	\begin{defn}[Non-crossing partition \cite{NS2006}]
		Given two elements $i, j \in [n],$ we write $i \sim_\pi j$ if $i$ and $j$ belongs to the same block of $\pi.$ A partition $\pi$ is called crossing if if there exist $i_1 < j_1 < i_2 < j_2 \in [n]$ such that $i_1 \sim_\pi i_2 \nsim_\pi j_1 \sim_\pi j_2.$ We call $\pi$ a non-crossing partition if $\pi$ is not crossing, and we denote $\mathcal{NC}(n)$ by the set of all non-crossing partitions of $[n].$
	\end{defn}

Let $\mathcal{S}_n$ be the symmetric group of permutations on $[n]$. For any permutation $\alpha\in\mathcal{S}_n$, let $\#(\alpha)$ be the number of cycles of the permutation $\alpha$, and $|\alpha|$ be the minimal number of transpositions needed to decompose $\alpha$.  Note that 
	\begin{equation}\label{length-cycle}
		|\alpha| = n-\#(\alpha), \text{ for any }\alpha\in \mathcal{S}_n.
	\end{equation}
	
	\begin{defn}[Non-crossing permutation \cite{NS2006}]	
		Let $\gamma_n$ be the full cycle $(1, 2, \ldots, n)$, and $\Id_n$ be the identity permutation, i.e., $\Id_n(i) = i$ for all $i\in[n]$. We call $\alpha \in \mathcal{S}_n$ a non-crossing permutation if $|\alpha|+|\alpha^{-1}\gamma_n| = n-1$, and we denote $\mathcal{S}_{\mathcal{NC}}(\gamma_n)$ by the set of all non-crossing permutations of $\mathcal{S}_n.$
	\end{defn}
	
	Similarly, there is also a partial order on $\mathcal{S}_{\mathcal{NC}}(\gamma_n)$. For $\alpha,\beta\in \mathcal{S}_{\mathcal{NC}}(\gamma_n)$,
	\begin{equation}\label{permutation-ineq}		
		\alpha\leq\beta \; \text{iff} \; |\alpha|+|\alpha^{-1}\beta|=|\beta|.
	\end{equation}
	Moreover, Biane showed that there is an isomorphism between $\mathcal{NC}(n)$ and $\mathcal{S}_{\mathcal{NC}}(\gamma_n)$, which preserves the partial order.  

\begin{defn}[\cite{MS17}]
		Let $\gamma_{n,n} = (1,\ldots,n)(n+1,\ldots,2n) \in \mathcal{S}_{2n}.$ We call $\alpha \in \mathcal{S}_{2n}$ a connected permutation if there exists at least one cycle of $\alpha$ which contains elements both from $\{1, \ldots, n\}$ and $\{n+1, \ldots, 2n\}.$ We denote
		$\mathcal{S}_{2n,c}$ by the set of all the connected permutations of $\mathcal{S}_{2n}$.
	\end{defn}
	

	\subsection{Weingarten's formula}
	
	Let $\mathbb{U}_N(\mathbb{C})$ be the set of all $N \times N$ unitary matrices. It is known that there is a (unique) Haar measure on $\mathbb{U}_N(\mathbb{C}),$ which induces a probability space structure for each entry of $U \in \mathbb{U}_N(\mathbb{C}).$ 
	
	\begin{thm}[Weingarten's formula \cite{NS2006}]\label{thm:Weingarten}
		Let $U=(u_{i,j})_{i,j=1}^N \in \mathbb{U}_N(\mathbb{C})$, we have for all $N \geq n$ that
		\begin{equation*}
			\begin{split}
			&\mathbb{E}\big[u_{i_1, j_1}\cdots u_{i_n,j_n}\bar{u}_{i_1^\prime, j_1^\prime}\cdots\bar{u}_{i_n^\prime, j_n^\prime}\big]\\
			&=\sum_{\alpha,\beta\in \mathcal{S}_n}\delta_{i_1i_{\beta(1)}^\prime}\cdots\delta_{i_ni_{\beta(n)}^\prime}\delta_{j_1j_{\alpha(1)}^\prime}\dots\delta_{j_nj_{\alpha(n)}^\prime}\Wg(N,\beta\alpha^{-1}),
			\end{split}
		\end{equation*}
		where we have used that 
		$$\Wg(N,\alpha) := \mathbb{E}\left[ u_{1,1}\cdots u_{n,n}\bar{u}_{1,\alpha(1)}\cdots \bar{u}_{n,\alpha(n)}\right].$$
		And $\Wg(\cdot, \cdot)$ is called the Weingarten function.
	\end{thm}
	
	\begin{rem}
		We will frequently use the following approximation equality for the Weingarten function \cite{BC03}. 		
		\begin{equation}\label{Wg function}
			\Wg(N,\alpha) = \phi(\alpha)N^{-|\alpha|-n}+O(N^{-|\alpha|-n-2}),
		\end{equation} 
		where $\phi(\alpha)$ is a function that is not depend on $N.$ 
	\end{rem}

	
	\subsection{Random graph states}\label{subsection-graph}

	Let $G=(V,E)$ be a simple undirected graph with the vertex $V=\{v_1,\ldots,v_k\}$ and the edge $E=\{e_1,\ldots, e_m\}$. We allow multiple edges between two vertices as well as vertex loops. For each $i\in[k]$, let $d_i$ be the degree of the vertex $v_i$. Hence, one has $\sum_{i=1}^{k}d_i = 2m$. 
		
	For a given graph $G$, one can associate a fattened graph $\mathcal{G} =(\mathcal{V}, \mathcal{E})$ as follows \cite{BIK3}: The vertex set $\mathcal{V}=\{\mathcal{V}_1, \ldots, \mathcal{V}_k\}$, where $\mathcal{V}_i$ is the set of $d_i$-copies of $v_i, i=1, \ldots, k.$ Thus, $\mathcal{V}$ can be equivalently considered as a partition of $[2m].$
	The edge set $\mathcal{E}=\{\mathcal{E}_1, \ldots, \mathcal{E}_m\}$ corresponds to the edges of $G$ in such a way that every two edges are now disjoint. Namely, $\mathcal{V}_i, \mathcal{V}_j \in \mathcal{E}_l$ for some $l \in [m]$ if and only if $v_i, v_j \in e_l.$ 
		
	Let $\mathcal{S}$ and $\mathcal{T}$ be two subsets of $[2m]$ such that $\mathcal{S}\cap \mathcal{T} = \emptyset$ and $\mathcal{S}\cup \mathcal{T} = [2n]$. Then $\{\mathcal{S},\mathcal{T}\}$ is a partition on $[2n]$. For any vertex $\mathcal{V}_i\in\mathcal{V}$, denote $\mathcal{S}_i = \mathcal{S}\cap \mathcal{V}_i$ and $\mathcal{T}_i =\mathcal{T}\cap \mathcal{V}_i$. Let $s_i: = \#(\mathcal{S}_i)$ and $t_i: = \#(\mathcal{T}_i)$ for all $1\leq i\leq k$, we denote 
 \begin{equation}\label{s-t-sequence}
		[s_i,t_i]_{i=1}^{k}:=\{\{\tilde{\mathcal{S}},\tilde{\mathcal{T}}\}: \tilde{\mathcal{S}}\subset[2n],\ \tilde{\mathcal{T}}\subset[2n],\ \tilde{\mathcal{S}}\cap \tilde{\mathcal{T}} = \emptyset \text{ and } \tilde{\mathcal{S}}\cup \tilde{\mathcal{T}} = [2n],\ \#(\tilde{\mathcal{S}}_i )= s_i,\ \# (\tilde{\mathcal{T}}_i)=t_i, i =1, \ldots, k\}.
	\end{equation}

	\begin{defn}[\cite{BIK3}]\label{def-area}
		$\{\mathcal{S}, \mathcal{T}\}$ is a partition on $[2n]$. An edge $\mathcal{E}_l$ in $\mathcal{G}$ is called crossing if $\mathcal{E}_l\cap \mathcal{S}\not=\emptyset\text{ and }\mathcal{E}_l\cap \mathcal{T}\not=\emptyset$. $[s_i,t_i]_{i=1}^{k}$ is the set of partitions on $[2n]$ given as above.
		For any $\{\tilde{\mathcal{S}},\tilde{\mathcal{T}}\}\in[s_i,t_i]_{i=1}^{k}$, we define the number of crossings by
		\begin{equation*}
		Cr(\{\tilde{\mathcal{S}},\tilde{\mathcal{T}}\})=\# \{l \in [m]: \mathcal{E}_l \; \text{is crossing}\}.
		\end{equation*}
		Moreover, the area of the boundary (or the area) of  $\{\mathcal{S},\mathcal{T}\}$ is defined by
		\begin{equation*}
		|\partial\mathcal{S}|=\max_{\{\tilde{\mathcal{S}},\tilde{\mathcal{T}}\}\in[s_i,t_i]_{i=1}^{k}}Cr(\{\tilde{\mathcal{S}},\tilde{\mathcal{T}}\}).
		\end{equation*}
	\end{defn}
	
	Now we are ready to introduce the random graph state. For a given graph $G$ and its fattened graph $\mathcal{G},$ the total quantum system is described by a tensor product of Hilbert space $\mathcal{H}=\bigotimes_{i=1}^{k} \left( \bigotimes_{j\in \mathcal{V}_i}\mathcal{H}_j \right)$, where $\mathcal{H}_j$ is an $N-$dimensional Hilbert space for each $j\in\mathcal{V}_i.$
	
\begin{defn}[Random graph state \cite{BIK0}]\label{def-uni}		
For any vertex $\mathcal{V}_i,$ one can associate a Haar unitary random matrix $U_i\in\mathbb{U}_{N^{d_i}}(\mathbb{C})$ acting on $\bigotimes_{j\in \mathcal{V}_i}\mathcal{H}_j,$
		$i =1, \ldots, k.$ And for any edge $\mathcal{E}_l=\{l_1,l_2\},$ let $|\psi_{l_1, l_2}\rangle$ be the maximumly entangled state acting on $\mathcal{H}_{l_1} \otimes \mathcal{H}_{l_2}.$ 
		Assume that $U_1,U_2,\cdots,U_k$ are independent, then the random graph state $|\Psi_\mathcal{G}\rangle$ is defined by 
		\begin{equation}\label{state-1}
			|\Psi_\mathcal{G}\rangle = \left[ \bigotimes_{1\leq i\leq k}U_i\right] \left(\bigotimes_{1\leq l\leq m}\big|\psi_{l_1, l_2}\big\rangle\right).
		\end{equation}
		Moreover, for given subsets $\mathcal{S}, \mathcal{T} \subseteq [2n],$ define the marginal of the graph state by
		\begin{equation}
			\rho_\mathcal{S} = \Tr_\mathcal{T}|\Psi_\mathcal{G}\rangle\langle\Psi_\mathcal{G}|,
		\end{equation}
		where $\Tr_\mathcal{T}$ is the partial trace over $\bigotimes_{i\in\mathcal{T}}\mathcal{H}_i.$
	\end{defn}

	
\subsection{Flow network and maximum flow problem}\label{subsec:flownetwork}

In this subsection, we recall the notions of flow network and the maximum flow problem. We refer to \cite{CL22} for more details.

\begin{defn}
		A flow network $\bfG  = (\bfV,\bfE)$ is a directed graph in which each edge $(\bfu,\bfv)\in\bfE$ has a non-negative capacity $\bfc(\bfu,\bfv)$.
	\end{defn}
	 If $(\bfu,\bfv)\not\in\bfE$, we define $\bfc(\bfu,\bfv)=0$, and we disallow self-loops. We designate two vertices in a flow network: a source $\bfs$ and a sink $\bft$. For convenience, we assume that each vertex lies on some path from the source to the sink. That is, for each vertex $\bfv\in \bfV$, the flow network contains a path $\bfs\sim\bfv\sim\bft$. The graph is therefore connected.

	\begin{defn}
	A flow in $\bfG $ is a function $\bff:\bfV\times\bfV\to\mathbb{N}$ that satisfies the
	following three properties:
	\begin{itemize}
		\item{Capacity constraint:} For all $\bfu,\bfv\in\bfV$, we require $\bff(\bfu,\bfv)\leq \bfc (\bfu,\bfv)$;
		\item{Skew symmetry:} For all $\bfu,\bfv\in\bfV$, $\bff(\bfu,\bfv) =- \bff(\bfv,\bfu)$;
		\item{Flow conservation:} For all $\bfu\in \bfV\backslash\{\bfs,\bft\}$, we require 
		\begin{equation*}
		\sum_{\bfv\in\bfV}\bff(\bfv,\bfu)=0.
		\end{equation*}
	\end{itemize}
	The value of a flow $\bff$ is defined as
\begin{equation*}
\left|\bff\right| = \sum_{\bfu\in \bfV}\bff(\bfs,\bfu).
\end{equation*}
	\end{defn}
The maximum-flow problem is to find a flow that has a maximum value for a given flow network $\bfG $. 
Given a flow network $\bfG$ and a flow $\bff$, the residual network $\bfG_\bff$ consists of edges with capacities representing how the flow on the edges of $\bfG $ can be changed. An edge of the flow network can admit additional flow equal to the edge's capacity minus the flow on that edge. If this value is positive, we include that edge in $\bfG_\bff$ with a "residual capacity" of $\bfc_\bff(\bfu,\bfv)=\bfc(\bfu,\bfv)-\bff(\bfu,\bfv)$. The only edges from $\bfG $ in $\bfG_\bff$ are those that can admit more flow; edges $(\bfu,\bfv)$ with flow equal to their capacity have $\bfc_\bff(\bfu,\bfv) = 0$ and are not in $\bfG_\bff$.

	\begin{defn}
Let  $\bfG $ be a flow network. An augmenting path $\bfp$ is a simple path from $\bfs$ to $\bft$ in the residual network $\bfG_\bff$.
	\end{defn}
	By the definition of the residual network, the flow on an edge $(\bfu,\bfv)$ of an augmenting path $\bfp$ can be increased by up to $\bfc_\bff(\bfu,\bfv)$ without violating the capacity constraint on either $(\bfu,\bfv)$ or $(\bfv,\bfu)$ in the original flow network $\bfG $.
	The maximum amount by which we can increase the flow on each edge in an augmenting path $\bfp$ is called the residual capacity of $\bfp$, defined as
	\[
	\bfc_\bff(\bfp) = \min\{\bfc_\bff(\bfu,\bfv):(\bfu,\bfv) \text{ is on } \bfp\}.
	\]
	
Solutions to the maximum flow problem are obtained as sums of augmenting paths, as per the Ford-Fulkerson algorithm \cite[Chapter 26]{CL22}. The process begins with an empty flow $\bff\equiv 0$. For an augmenting path $\bfp$ with residual capacity $\bfx$, the flow function is updated by adding $\bfx$ units of flow to each edge $(\bfu,\bfv)\in \bfp$. The residual capacities of edges in the new residual network are updated by subtracting $\bfx$ from the capacities of edges in the previous residual network. This process is iterated until no more augmenting paths exist. The maximum value of flow that can be sent from $\bfs$ to $\bft$ is denoted as $\bfX= \left|\bff\right|$. For a fixed flow network, there may be many flows such that their value is the maximum value of flow that can be sent in this flow network.
	
Now we turn to consider the relation between the maximal flow and the area of  $\{\mathcal{S},\mathcal{T}\}.$ Recall that $t_i, s_i, $ and $d_i$ are given in Subsection \ref{subsection-graph}. Let $\bet_1,\ldots,\bet_k$ be in $\mathcal{S}_n$. Define a flow network $\tilde{\bfG}$ as follows:
The vertex set is given by $\tilde{\bfV} = \{\Id_n,\gamma_n,\bet_1,\ldots,\bet_k\},$ with two distinguished vertices: source $\bfs = \gamma_{n}$ and the sink $\bft = \Id_n.$ The edges in $\tilde{\bfE}$ are oriented and they are of three types:
\begin{equation*}
\tilde{\bfE} = \{(\bet_{i},\Id_n): t_i>0 \}\cup\{(\gamma_{n},\bet_{i}):s_i>0\}\cup\{(\bet_{i},\bet_{j}), (\bet_{j},\bet_{i}):e_{i,j}>0,\ i<j\}.
\end{equation*}
The capacities of the edges are given by
\begin{equation*}
\bfc(\bet_{i},\Id_n) = t_i>0,\ \bfc(\gamma_{n},\bet_{i}) = s_i>0,\ \bfc(\bet_{i},\bet_{j}) = \bfc(\bet_{j},\bet_{i}) = e_{i,j}>0.
\end{equation*}
Let $\tilde{\bfX}$ be the maximal value of flow in network $\tilde{\bfG}$ can be sent from source $\gamma_{n}$ to sink $\Id_{n}$. 
We conclude this subsection with the following theorem:
		
\begin{thm}[Theorem 5.2 \cite{BIK3}]\label{flow-area}
			Let $\tilde{\bfX}$ be the maximal value of the flow that can be sent from the source to the sink in $\tilde{\bfG},$ and $|\partial\mathcal{S}|$ be the area of boundary of $\{\mathcal{S},\mathcal{T}\}$ in $\mathcal{G}$. Then we have
\begin{equation*}
\tilde{\bfX} = |\partial\mathcal{S}|.
\end{equation*}
\end{thm}
	


\section{Main results}

In \cite{BIK3}, it was demonstrated that the entropy of entanglement in random graph states follows an average area law, i.e., $\mathbb{E}(H(\rho_S))\propto|\partial\mathcal{S}|$. More precisely, for a random graph state $\rho_{\mathcal{S}}$ given in Definition \ref{def-uni}, denote $\mu_N$ by the empirical eigenvalue measure of $N^{\tilde{\bfX}} \rho_{\mathcal{S}}$, i.e.,
\begin{equation}
		\mu_{N} = \frac{1}{N^{\tilde{\bfX} }}\sum_{i=1}^{N^{\tilde{\bfX}}}\delta_{N^{\tilde{\bfX} }\lambda_i(\rho_{\mathcal{S}})}.
\end{equation}
Then, we have for any $n \geq 1,$
\begin{equation}
\mathbb{E} \int_{\mathbb{R}} t^n d \mu_{N} = \mathbb{E}\left[\frac{1}{N^{\tilde{\bfX}} }\Tr\left[N^{\tilde{\bfX}} \rho_{\mathcal{S}}\right]^n\right].
\end{equation}
It was shown in \cite{BIK3} that there exists a compactly supported probability measure $\mu$ on $\mathbb{R}$, such that $\mathbb{E} \mu_N$ converges weakly to $\mu,$ as 
$N \rightarrow \infty.$ Consequently, we have

\begin{thm}[Theorem 6.1 \cite{BIK3}]\label{unitary area law}
		Let $\rho_\mathcal{S}$ be the marginal $\{\mathcal{S}, \mathcal{T}\}$ of a graph state $\mathcal{G}$. Then, as $N\to\infty$, the area law holds, in the following sense
\begin{equation*}
\mathbb{E}[H(\rho_\mathcal{S})]=|\partial\mathcal{S}|\log N- \int_{\mathbb{R}} t\log td\mu +o(1),
\end{equation*}
where $|\partial\mathcal{S}|$ is the area of the boundary of the partition $\{\mathcal{S}, \mathcal{T}\},$ and $H (\rho_{\mathcal{S}}): =- {\rm Tr} [\rho_{\mathcal{S}} \log \rho_{\mathcal{S}}]$ is the von Neumann entropy of $\rho_{\mathcal{S}}.$  
\end{thm}
The key step for this result is to show that $\mathbb{E} \mu_N$ converges in moments to $\mu$ as $N \rightarrow \infty.$ Let $\{\nu_N, \nu\}$ be a sequence of probability measures on $\mathbb{R}$, we say that $\nu_N \rightarrow \nu$ in moments as $N \rightarrow \infty,$ if
\begin{equation*}
\int_{\mathbb{R}} t^n d\nu_N  \rightarrow \int_{\mathbb{R}} t^n d\nu
\end{equation*}
for all $n \geq 1.$ We note that $\mathbb{E} \nu_N \rightarrow \nu$ in moments if 
\begin{equation*}
\mathbb{E} \int_{\mathbb{R}} t^n d\nu_N  \rightarrow \int_{\mathbb{R}} t^n d\nu
\end{equation*}
for all $n \geq 1.$ In this paper, our goal is to enhance the above results by establishing the almost sure convergence. Namely, we shall prove the following theorem.
 
\begin{thm}\label{thm:almost-sure}
The convergence in Theorem \ref{unitary area law} holds almost surely, i.e., almost surely as $N \rightarrow \infty,$
\begin{equation*}
H(\rho_\mathcal{S})=|\partial\mathcal{S}|\log N- \int_{\mathbb{R}} t\log td\mu +o(1).
\end{equation*}
\end{thm}
The key step for proving this theorem is the following.
\begin{prop}\label{prop:almost surely}
Almost surely as $N \rightarrow \infty$, $\mu_{N}$ converges in moments to $\mu$.
\end{prop}
Since $\mu$ is compactly supported, it is sufficient to show that 
\begin{equation}\label{eq:covariance}
\mathbb{E}\bigg[\bigg(\int_{\mathbb{R}} t^n \mathrm{d}\mu_{N}\bigg)^2\bigg]-\bigg[\mathbb{E}\bigg(\int_{\mathbb{R}} t^n \mathrm{d}\mu_{N}\bigg)\bigg]^2 = O(N^{-2})
\end{equation}
for any $n\geq 1$. Using the Weingarten calculus, we treat the expectations separately. For the first one, we can write
			\begin{equation}\label{eq:uni-cov-1}
				\begin{split}
				& \mathbb{E}\bigg[\bigg(\frac{1}{N^{\tilde{\bfX}} }\Tr\big[N^{\tilde{\bfX}} \rho_\mathcal{S}\big]^n\bigg)^2\bigg]\\
				& = \sum_{\substack{\alpha_1,\cdots,\alpha_k,\\\beta_1,\cdots,\beta_k\in\mathcal{S}_{2n}}}N^{\tilde{\bfX}(2n-2)}\prod_{i=1}^{k}N^{\#(\gamma_{n,n}^{-1}\alpha_{i}) s_i}\prod_{i=1}^{k}N^{\#(\alpha_{i})t_i}\prod_{1\leq i\leq j\leq k}N^{\#(\beta_i\beta_j^{-1})e_{i,j}}\\
				& \quad \times \prod_{1\leq i\leq j\leq k}N^{-2ne_{i,j}}\prod_{i=1}^{k}\Wg(N^{d_i},\beta_{i}\alpha_i^{-1}).
			\end{split}
			\end{equation}
We have used the following fact \cite[Theorem 5.1]{BIK0}: for each independent unitary matrix $U_i$, $N^{\#(\alpha_{i})t_i}$ counts the contribution of the loops corresponding to partial traces, $N^{\#(\gamma_{n,n}^{-1}\alpha_{i}) s_i}$ corresponds to the moment product, $N^{\#(\beta_i\beta_j^{-1})e_{i,j}}$ represents the contribution of the loops coming from the Bell states	between vertices $\mathcal{V}_i$ and $\mathcal{V}_j$. It follows that 
			\begin{equation}\label{eq:uni-cov-1.1}
			\begin{split}
				 \eqref{eq:uni-cov-1}& = \sum_{\substack{\alpha_1,\cdots,\alpha_k,\\\beta_1,\cdots,\beta_k\in\mathcal{S}_{2n}}}N^{\tilde{\bfX}(2n-2)}\prod_{i=1}^{k}N^{\#(\gamma_{n,n}^{-1}\alpha_{i}) s_i}\prod_{i=1}^{k}N^{\#(\alpha_{i})t_i}\prod_{1\leq i\leq j\leq k}N^{\#(\beta_i\beta_j^{-1})e_{i,j}}\\
				 & \quad \times\prod_{1\leq i\leq j\leq k}N^{-2ne_{i,j}}\prod_{i=1}^{k}\bigg(\phi(\beta_i\alpha_i^{-1})N^{(\#(\beta_{i}\alpha_i^{-1})-4n)d_i}+O(N^{-2})\bigg)\\
				& = \sum_{\substack{\alpha_1,\cdots,\alpha_k,\\\beta_1,\cdots,\beta_k\in\mathcal{S}_{2n}}}N^{\tilde{\bfX}(2n-2)}\prod_{i=1}^{k}N^{2nd_i-|\gamma_{n,n}^{-1}\alpha_{i}| s_i-|\alpha_{i}|t_i}\prod_{1\leq i<j\leq k}N^{-|\beta_i\beta_j^{-1}|e_{i,j}}\\
				& \quad \times \prod_{i=1}^{k}N^{(-|\beta_{i}\alpha_i^{-1}|-2n)d_i}\bigg(\prod_{i=1}^{k}\phi(\beta_i\alpha_i^{-1})+O(N^{-2})\bigg)\\
				& = \sum_{\substack{\alpha_1,\cdots,\alpha_k,\\\beta_1,\cdots,\beta_k\in\mathcal{S}_{2n}}}N^{\tilde{\bfX} (2n-2)-F_{n,n}(\alpha,\beta)}\bigg(\prod_{i=1}^{k}\phi(\beta_i\alpha_i^{-1})+O(N^{-2})\bigg),
				\end{split}
			\end{equation}
where we denote 
\begin{equation}
F_{n,n}(\alpha,\beta) := \sum_{i=1}^{k}\big(s_i|\gamma_{n,n}^{-1}\alpha_i| + t_i|\alpha_i| \big)+\sum_{1\leq i < j\leq k}e_{i,j}|\beta_i\beta_j^{-1}|+\sum_{i=1}^{k} d_i|\beta_i\alpha_i^{-1}|.
\end{equation}

For the second expectation, we have
\begin{equation}\label{eq:uni-cov-2}
			\begin{split}
				& \bigg[\frac{1}{N^{\tilde{\bfX}} }\mathbb{E}\bigg(\Tr\big[N^{\tilde{\bfX}} \rho_\mathcal{S}\big]^n\bigg)\bigg]^2\\
				& = N^{\tilde{\bfX}(2n-2)} \bigg[\sum_{\substack{\alpha_1,\cdots,\alpha_k,\\\beta_1,\cdots,\beta_k,\in\mathcal{S}_n}} \prod_{i=1}^{k}N^{\#(\gamma_n^{-1}\alpha_{i}) s_i}\prod_{i=1}^{k}N^{\#(\alpha_{i})t_i}\prod_{1\leq i\leq j\leq k}N^{\#(\beta_i\beta_j^{-1})e_{i,j}}\\
				&\quad\times \prod_{1\leq i\leq j\leq k}N^{-ne_{i,j}}\prod_{i=1}^{k}\Wg(N^{d_i},\beta_{i}\alpha_i^{-1})\bigg]^2\\
				& = N^{\tilde{\bfX}(2n-2)} \bigg[\sum_{\substack{\alpha_1,\cdots,\alpha_k,\\\beta_1,\cdots,\beta_k,\in\mathcal{S}_n}} \prod_{i=1}^{k}N^{nd_i-|\gamma_n^{-1}\alpha_{i}| s_i-|\alpha_{i}|t_i}\prod_{1\leq i<j\leq k}N^{-|\beta_i\beta_j^{-1}|e_{i,j}}\\
				& \quad \times \prod_{i=1}^{k}\bigg(\phi(\beta_{i}\alpha_i^{-1})N^{d_i(|-\beta_i\alpha_i^{-1}|-n)}+O(N^{-2})\bigg)\bigg]^2\\
				& = \sum_{\substack{\alpha_1^{(1)},\cdots,\alpha_k^{(1)},\beta_1^{(1)},\cdots,\beta_k^{(1)},\\ \alpha_1^{(2)},\cdots,\alpha_k^{(2)}, \beta_1^{(2)},\cdots, \beta_k^{(2)}\in\mathcal{S}_n}}N^{\tilde{\bfX} (2n-2)-F_n(\alpha^{(1)},\beta^{(1)})-F_n(\alpha^{(2)},\beta^{(2)})}\\
				& \quad \times \bigg(\prod_{i=1}^{k}\phi(\beta_i^{(1)}(\alpha_i^{(1)})^{-1})\phi(\beta_i^{(2)}(\alpha_i^{(2)})^{-1})+O(N^{-2})\bigg),
			\end{split}
			\end{equation}
where we denote 
\begin{equation}
F_n(\alpha^{(s)},\beta^{(s)}) :=\sum_{i=1}^{k}\big(s_i|\gamma_{n}^{-1}\alpha_i^{(s)}| + t_i|\alpha_i^{(s)}| \big)+\sum_{1\leq i < j\leq k}e_{i,j}|\beta_i^{(s)} (\beta_j^{(s)})^{-1}|+\sum_{i=1}^{k}d_i|\beta_i^{(s)}(\alpha_i^{(s)})^{-1}| 
\end{equation}
$s=1, 2.$

Hence, to prove \eqref{eq:covariance}, we need to consider the difference between the dominant terms in \eqref{eq:uni-cov-1} and \eqref{eq:uni-cov-2}. This motivates us to make the following definition:

\begin{defn}
{\rm (1).} Let $\{\alpha_{i}\}_{i=1}^k$ and $\{\beta_{i}\}_{i=1}^k$ be two sequences of permutations in $\mathcal{S}_{2n}$. We define the function $F_{n,n}(\cdot,\cdot)$ as:
		\begin{equation*}
			\begin{split}
F_{n,n}(\{\alpha_{i}\}_{i=1}^k,\{\beta_{i}\}_{i=1}^k) := & \sum_{i=1}^{k} \left(s_i|\gamma_{n,n}^{-1}\alpha_i|+t_i|\alpha_i|\right) + \sum_{1\leq i < j\leq k}e_{i,j}|\beta_i^{-1}\beta_j| + \sum_{i=1}^{k}d_i|\beta_i\alpha_i^{-1}|.
			\end{split}
		\end{equation*}
{\rm (2)}.  Let $\{\alpt_i\}_{i=1}^k$ and $\{ \bet_i \}_{i=1}^k$ be two sequences of permutations in $\mathcal{S}_n$. We define 
				\begin{equation*}
					\begin{split}
						F_n(\{\alpt_{i}\}_{i=1}^k,\{\bet_{i}\}_{i=1}^k):=\sum_{i=1}^{k}(s_i|\gamma_n^{-1}\alpt_i|+t_i|\alpt_i|)+\sum_{1\leq i < j\leq k}e_{i,j}|\bet_i^{-1}\bet_j|+\sum_{i=1}^{k}d_i|\bet_i\alpt_i^{-1}|.
					\end{split}
				\end{equation*}
For simplicity, we denote $F_{n,n} (\alpha, \beta):= F_{n,n}(\{\alpha_{i}\}_{i=1}^k,\{\beta_{i}\}_{i=1}^k)$ and $F_{n} (\alpt, \bet):= F_{n}(\{\alpt_{i}\}_{i=1}^k,\{\bet_{i}\}_{i=1}^k).$
\end{defn}
Thus, the dominant terms of \eqref{eq:uni-cov-1} and \eqref{eq:uni-cov-2} are determined by the minimal values of $F_{n,n} (\alpha, \beta)$ and $F_{n} (\alpt, \bet).$ We postpone the minimization problem of  $F_{n,n} (\alpha, \beta)$ and $F_{n} (\alpt, \bet)$ to the next subsection.


\subsection{Minimization problem of $F_{n,n} (\alpha, \beta)$ and $F_{n} (\alpt, \bet)$}\label{subsec:min}

We note that the minimization problem of $F_{n} (\alpt, \bet)$ is considered in \cite[Section 5.2]{BIK0}. The minimization problem for $F_{n} (\alpt, \bet)$ can be translated into the maximum flow problem on the network $\tilde{\bfG}  = (\tilde{\bfV}, \tilde{\bfE})$ given in Subsection \ref{subsec:flownetwork}. Namely, we have the following proposition:

\begin{prop}[\cite{BIK0}]\label{Fn min cond}
For any $\{\alpt_i\}_{i=1}^k$ and $\{ \bet_i \}_{i=1}^k$ in $\mathcal{S}_n,$
\begin{equation}
F_n(\alpt,\bet) \geq \tilde{\bfX} (n-1).
\end{equation}
The equality holds if and only if, there exists a maximal flow $\tilde{\bff}$ in $\tilde{\bfG}$ with $|\tilde{\bff}| = \tilde{\bfX}$, satisfying the following statements:
			\begin{enumerate}[{\rm (I)}]
				\item For any augmenting path 
				$$\tilde{\bfp}_i:\gamma_{n}\to \bet_{i_{1}}\to\bet_{i_{2}}\to\cdots\to\bet_{i_{l(i)}}\to\Id_{n},$$
				we have $\bet_{i_{l(i)}}\leq \cdots\leq\bet_{i_2}\leq \bet_{i_l}\leq \gamma_{n}$ for $1\leq i\leq l$;\label{Fn-min-1}
				\item For all $1\leq i\leq k,$ $\bet_{i} = 
				\begin{cases}
					\Id_{n},& \text{if } \bfc_{\tilde{\bff}}(\bet_{i},\Id_n)>0;\\
					\gamma_{n},& \text{if } \bfc_{\tilde{\bff}}(\gamma_n,\bet_i)>0;\\ 
					\bet_{j},& \text{if there exist } 1\leq j\leq k \text{ such that } \bfc_{\tilde{\bff}}(\bet_i,\bet_j)>0;
				\end{cases}$\label{Fn-min-2}
				\item For all $1\leq i\leq k,$ $\alpt_{i}$ satisfies $
				\begin{cases}
					\alpt_i = \bet_i,& \text{if } s_i > 0 \text{ and } t_i > 0;\\
					\Id_{n}\leq \alpt_i \leq \bet_i,& \text{if } s_i = 0\text{ and } t_i > 0;\\ 
					\bet_i\leq \alpt_i\leq \gamma_{n},& \text{if } s_i > 0 \text{ and } t_i = 0.
				\end{cases} $\label{Fn-min-3}
			\end{enumerate}
\end{prop}

Motivated by the above proposition, we can study the minimization problem of $F_{n,n} (\alpha, \beta).$ Firstly, we introduce the following definition:
\begin{defn}\label{def-Fnn}
Let $\{\beta_{i}\}_{i=1}^k$ be a sequence of permutations in $\mathcal{S}_{2n}$. We define
\begin{equation*}
			F_{n,n}(\{\beta_{i}\}_{i=1}^k) = \sum_{i=1}^{k} \left(s_i|\gamma_{n,n}^{-1}\beta_i|+t_i|\beta_i|\right) + \sum_{1\leq i < j\leq k}e_{i,j}|\beta_i^{-1}\beta_j|.
		\end{equation*}
		We will use $F_{n,n}(\beta)$ to denote $F_{n,n}(\{\beta_{i}\}_{i=1}^k)$.
\end{defn}
Similarly, we can associate the minimization problem for $F_{n,n}(\alpha, \beta)$ to the maximum flow problem on a network $\bfG  = (\bfV,\bfE)$, which is given as follows:
The vertex set is given by $\bfV = \{\Id_{2n},\gamma_{n,n},\beta_1,\ldots,\beta_k\},$ with two distinguished vertices: the source $\bfs = \gamma_{n,n}$ and the sink $\bft = \Id_{2n}.$ The edges in $\bfE$ are oriented and they are of three types:
\begin{equation*}
\bfE = \{(\beta_{i},\Id_{2n}): t_i>0 \}\cup\{(\gamma_{n,n},\beta_{i}):s_i>0\}\cup\{(\beta_{i},\beta_{j}), (\beta_{j},\beta_{i}):e_{i,j}>0,\ i< j\}.
\end{equation*}
The capacities of the edges are given by
\begin{equation*}
\bfc(\beta_{i},\Id_{2n}) = t_i>0,\ \bfc(\gamma_{n,n},\beta_i) = s_i>0,\ \bfc(\beta_{i},\beta_{j}) = \bfc(\beta_{j},\beta_{i}) = e_{i,j}>0.
\end{equation*}	

Finding the maximal flow in $\bfG$ is standard, based on the Ford-Fulkerson algorithm. We note that the following process is similar to the process for finding the maximal flow in $\tilde{\bfG}$; see \cite[Section 5.2]{BIK0}. However, we make some necessary modifications for completeness. Let us begin with an empty flow $\bff_0 = 0$. For any augmenting path $\bfp_1$ in the residual network $\bfG_{\bff_0}$ with residual capacity $\bfX_1$,
\begin{equation*}
	\bfp_1 : \gamma_{n,n}\to \beta_{i_{1}} \to \beta_{i_{2}} \to \cdots \to \beta_{i_{l(1)}} \to \Id_{2n}.
\end{equation*}
	One can use the inequality
\begin{equation*}
	\bfX _1 \left[|\gamma_{n,n}^{-1}\beta_{i_{1}}| + |\beta_{i_1}^{-1}\beta_{i_2}| + \cdots + |\beta_{i_{l(1)-1}}^{-1}\beta_{i_{l(1)}}| + |\beta_{i_{l(1)}}|\right] \geq \bfX _1(2n-2)
\end{equation*}
	in the function $F_{n,n}(\beta)$, 
and denote the remaining part of the function $F_{n,n}(\beta)$ by $F_{n,n}^{(1)}(\beta)$, i.e.,
\begin{equation*}
F_{n,n}^{(1)}(\beta) := F_{n,n}(\beta) - \bfX_1\left[|\gamma_{n,n}^{-1}\beta_{i_{1}}| + |\beta_{i_1}^{-1}\beta_{i_2}| + \cdots + |\beta_{i_{l(1)-1}}^{-1}\beta_{i_{l(1)}}| + |\beta_{i_{l(1)}}|\right].
\end{equation*}
	Hence, we have
\begin{equation*}
	F_{n,n}(\beta) \geq \bfX _1(2n-2) + F_{n,n}^{(1)}(\beta).
\end{equation*}
We can update the flow $\bff_0$ to the flow $\bff_1$, by adding $\bfX_1$ units of flow to each edge in $\bfp_1.$ The capacities of edges in the residual network $\bfG_{\bff_1}$, denoted by $\bfc_{\bff_1},$ are equal to the coefficients in $F_{n,n}^{(1)}(\beta)$. Specifically, we mean
	\begin{equation}
		\begin{cases}
			\bfc_{\bff_1}(\beta_i,\beta_j) = \bfc_{\bff_1}(\beta_j,\beta_i) = e_{i,j} - \bfX _1 & \text{if } 1 \leq i<j \leq k; \\
			\bfc_{\bff_1}(\gamma_{n,n},\beta_i) =  s_i - \bfX _1 & \text{if } 1 \leq i \leq k;\\
			\bfc_{\bff_1}(\Id_{2n},\beta_j)  = t_j - \bfX _1 & \text{if } 1 \leq j \leq k.
		\end{cases}
	\end{equation}
	The right part also be the coefficients in $F_{n,n}^{(1)}(\beta)$.
	
Next, 
we consider the augmenting path $\bfp_2$ in the residual network $\bfG _{\bff_1}$ with residual capacity $\bfX_2$,
\begin{equation*}
	\bfp_2 : \gamma_{n,n}\to \beta_{j_{1}} \to \beta_{j_{2}} \to \cdots \to \beta_{j_{l(2)}} \to \Id_{2n}.
\end{equation*}
	One can use the inequality
\begin{equation*}
	\bfX_2 \left[|\gamma_{n,n}^{-1}\beta_{j_{1}}| + |\beta_{j_1}^{-1}\beta_{j_2}| + \cdots +|\beta_{j_{l(2)-1}}^{-1}\beta_{j_{l(2)}}| + |\beta_{j_{l(2)}}|\right] \geq \bfX_2(2n-2)
\end{equation*}
	in the function $F_{n,n}^{(1)}(\beta)$, and denote the remaining part of the function $F_{n,n}^{(1)}(\beta)$ by $F_{n,n}^{(2)}(\beta)$, i.e.,
\begin{equation*}
F_{n,n}^{(2)}(\beta) := F_{n,n}^{(1)}(\beta) - \bfX_2 \left[|\gamma_{n,n}^{-1}\beta_{j_{1}}| + |\beta_{j_1}^{-1}\beta_{j_2}| + \cdots +|\beta_{j_{l(2)-1}}^{-1}\beta_{j_{l(2)}}| + |\beta_{j_{l(2)}}|\right].
\end{equation*}
	Hence, we have
\begin{equation*}
	F_{n,n}(\beta) \geq \bfX_1(2n-2) + F_{n,n}^{(1)}(\beta) \geq \bfX_1(2n-2) + \bfX_2(2n-2) + F_{n,n}^{(2)}(\beta).
\end{equation*}

We can update the flow $\bff_1$ to the flow $\bff_2$, by adding $\bfX_2$ units of flow to each edge in $\bfp_2.$ The capacities of edges in the residual network $\bfG_{\bff_2}$, denoted by $\bfc_{\bff_2}$, are equal to the coefficients in $F_{n,n}^{(2)}(\beta)$. Specifically, we mean
	\begin{equation}
		\begin{cases}
			\bfc_{\bff_2}(\beta_i,\beta_j) = \bfc_{\bff_2}(\beta_j,\beta_i) = \bfc_{\bff_1}(\beta_i,\beta_j) - \bfX_2 & \text{if } 1 \leq i<j \leq k; \\
			\bfc_{\bff_2}(\gamma_{n,n},\beta_i) =  \bfc_{\bff_1}(\gamma_{n,n},\beta_i) - \bfX_2 & \text{if } \ 1 \leq i \leq k;\\
			\bfc_{\bff_2}(\Id_{2n},\beta_j)  = \bfc_{\bff_1}(\Id_{2n},\beta_j) - \bfX_2 & \text{if } \ 1 \leq j \leq k.
		\end{cases}
	\end{equation}
	The right part also be the coefficients in $F_{n,n}^{(2)}(\beta)$.

Repeating the above process, updating the flow $\bff_{i-1}$ to $\bff_{i}$ by adding $\bfX_i$ units of flow to each edge in $\bfp_i$, until no augmenting path can be found in the residual network. By induction, we obtain the maximal flow $\bff$. Suppose we do this process $l$ times to obtain $\bff$, then $\bff$ is the superposition of all augmenting paths $\bfp_i$ with residual capacity $\bfX_i, 1\leq i\leq l$. Denote 
\begin{equation*}
\bff\simeq\{(\bfp_i,\bfX_i)\}_{i=1}^l.
\end{equation*}
The residual network of $\bfG$ induced by the maximum flow $\bff$ is $\bfG_\bff$ with the vertex $\bfV$, and the edges  
	\begin{equation*}
		\begin{split}
			\bfE_\bff=&\{(\beta_i,\Id_{2n})|\bfc_\bff(\Id,\beta_{i}) > 0, 1 \leq i \leq k \}\cup\{(\gamma_{n,n},\beta_i)|\bfc_\bff(\gamma_{n,n},\beta_i) > 0, 1 \leq i \leq k \}\\&\cup\{(\beta_i,\beta_j),(\beta_j,\beta_i)|\bfc_\bff(\beta_{i},\beta_j) = \bfc_\bff(\beta_j,\beta_i) > 0, 1 \leq i<j \leq k \}.
		\end{split}
	\end{equation*}
	Let $\bfX = |\bff|$ be the value of flow $\bff$. We have the following inequality:
	\begin{equation}\label{ine-res}
		\begin{split}
			F_{n,n}(\beta) \geq&\ \bfX _1(2n-2) + F_{n,n}^{(1)}(\beta)\\
			\geq&\ \bfX _1(2n-2) + \bfX _2(2n-2) + F_{n,n}^{(2)}(\beta)\\
			\cdots\\
			\geq&\ \bfX (2n-2) + \sum_{(\beta_i,\Id_{2n})\in\bfE_{\bff}}\bfc_\bff(\beta_{i},\Id_{2n})|\beta_i|\\
			&\ +\sum_{(\gamma_{n,n},\beta_i)\in\bfE_{\bff}}\bfc_\bff(\gamma_{n,n},\beta_i)|\gamma_{n,n}^{-1}\beta_i|+\sum_{(\beta_i,\beta_j)\in\bfE_{\bff},i<j}\bfc_\bff(\beta_i,\beta_j)|\beta_i^{-1}\beta_j|.
		\end{split}
	\end{equation}

In summary, we obtain the following solution for the minimization problem of $F_{n,n}(\alpha,\beta)$, which is an adaption of Proposition \ref{Fn min cond}.
\begin{prop}\label{Fnn min cond}
For any $\{\alpha_i\}_{i=1}^k$ and $\{ \beta_i \}_{i=1}^k$ in $\mathcal{S}_{2n},$
\begin{equation}\label{eq:inequality}
F_{n,n} (\alpha,\beta) \geq F_{n,n} (\beta) \geq \bfX (2n-2).
\end{equation}
The equality holds if and only if, there exists a maximal flow $\bff\simeq\{(\bfp_i,\bfX_i)\}_{i=1}^l$ in ${\bf G}$ with $|{\bf f}| = {\bf X}$, such that the following statements hold:
		\begin{enumerate}[{\rm (I)}]
			\item For any augmenting paths $\bfp_i$,
			 \begin{equation*}
			 \bfp_i:\gamma_{n,n}\to \beta_{i_{1}}\to\beta_{i_{2}}\to\cdots\to\beta_{i_{l(i)}}\to\Id_{2n},
			 \end{equation*}
			 we have $\beta_{i_{l(i)}}\leq \cdots\leq \beta_{i_2}\leq \beta_{i_1}\leq \gamma_{n,n}$, for all $1\leq i\leq l$;\label{Fnn-min-1}
			\item For all $1\leq i\leq k,$ $\beta_i = 
			\begin{cases}
				\Id_{2n},& \text{if } \bfc_\bff(\beta_{i},\Id_{2n})>0;\\
				\gamma_{n,n},& \text{if } \bfc_\bff(\gamma_{n,n},\beta_i)>0;\\ 
				\beta_{j},& \text{if there exist } 1\leq j\leq k,\ j\not= i, \text{ such that } \bfc_\bff(\beta_{i},\beta_{j})>0;
			\end{cases}$\label{Fnn-min-2}
			\item For all $1\leq i\leq k,$ $\alpha_i$ satisfies $
			\begin{cases}
				\alpha_i = \beta_i,& \text{if } s_i>0 \text{ and }t_i>0;\\
				\Id_{2n}\leq \alpha_i \leq \beta_i,& \text{if } s_i=0 \text{ and } t_i>0;\\
				\beta_i\leq \alpha_i\leq \gamma_{n,n},& \text{if } s_i>0 \text{ and } t_i=0.
			\end{cases}$\label{Fnn-min-3}
		\end{enumerate}
	\end{prop}

\begin{proof}
		Let us start from the function $F_{n,n}(\alpha,\beta)$. For any $1\leq i\leq k$, if $s_i > 0$ and $t_i > 0$, 
		\begin{equation*}
		|\gamma_{n,n}^{-1}\alpha_{i}|+|\beta_i\alpha_{i}^{-1}| \geq |\gamma_{n,n}^{-1}\beta_i| \text{ and } |\alpha_{i}|+|\beta_i\alpha_{i}^{-1}| \geq |\beta_i|;
		\end{equation*}
		if $s_i = 0$ and $t_i > 0$, we use inequality 
		\begin{equation*}|\alpha_{i}|+|\beta_i\alpha_{i}^{-1}| \geq |\beta_i|;
		\end{equation*}
		if $s_i > 0$ and $t_i = 0$, similarly we use inequality 
		\begin{equation*}|\gamma_{n,n}^{-1}\alpha_{i}|+|\beta_i\alpha_{i}^{-1}| \geq |\gamma_{n,n}^{-1}\beta_i|.
		\end{equation*}
		Combining with inequality \eqref{ine-res}, the inequality $F_{n,n}(\alpha,\beta)\geq F_{n,n}(\beta) \geq \bfX (2n-2)$ holds.

Now we turn to consider conditions for the equality of \eqref{eq:inequality}. Through the process of obtaining the inequality \eqref{ine-res}, the inequality $F_{n,n}(\beta)\geq\bfX (2n-2)$ takes equal if and only if, there exist one flow $\bff\simeq\{(\bfp_i,\bfX_i)\}_{i=1}^l$ satisfies $|\bff| = \bfX$, such that for any augmenting path $\bfp_i$ in $\bff$
\begin{equation*}
\bfp_i:\gamma_{n,n}\to \beta_{i_{1}}\to\beta_{i_{2}}\to\cdots\to\beta_{i_{l(i)}}\to\Id_{2n},
\end{equation*}
		with residual capacity $\bfX_i$, the inequality
\begin{equation*}
\bfX_i \big[|\gamma_{n,n}^{-1}\beta_{i_{1}}|+|\beta_{i_1}^{-1}\beta_{i_2}|+\cdots+|\beta_{i_{l(i)-1}}^{-1}\beta_{i_{l(i)}}|+|\beta_{i_{l(i)}}|\big]\geq \bfX_i (2n-2)
\end{equation*}
takes equal, and all the other terms from the residual network in the right hand of \eqref{ine-res} have no contribution. This implies Condition \eqref{Fnn-min-1} and Condition \eqref{Fnn-min-2}. Therefore, $F_{n,n}(\beta)=\bfX (2n-2)$ implies all the permutations $\{\beta_i\}_{i=1}^k$ lie on the geodesic path from $\Id_{2n}$ to $\gamma_{n,n}$.

Let us explore under what conditions the equality for $F_{n,n}(\alpha,\beta)\geq F_{n,n}(\beta)$ will be hold. If $s_i > 0$ and $t_i > 0$ for $1\leq i\leq k$, the solution to the equations
\begin{equation*}
|\gamma_{n,n}^{-1}\alpha_{i}|+|\beta_i\alpha_{i}^{-1}| = |\gamma_{n,n}^{-1}\beta_i| \; \text{and} \;  |\alpha_{i}|+|\beta_i\alpha_{i}^{-1}| = |\beta_i|
\end{equation*}
is $\alpha_i = \beta_i$. It is due to the following inequality.
\begin{equation*}
2n-2+2|\beta_i\alpha_{i}^{-1}|\leq |\gamma_{n,n}^{-1}\alpha_{i}|+|\beta_i\alpha_{i}^{-1}|+|\alpha_{i}|+|\beta_i\alpha_{i}^{-1}|=|\gamma_{n,n}^{-1}\beta_i|+|\beta_i|\leq 2n-2.
\end{equation*}
If $s_i = 0$ and $t_i > 0$, the condition $|\alpha_{i}|+|\beta_i\alpha_{i}^{-1}| = |\beta_i|$ implies that $\Id_{2n}\leq \alpha_{i}\leq \beta_i$. On the other hand, if $s_i > 0$ and $t_i = 0$, the condition $|\gamma_{n,n}^{-1}\alpha_{i}|+|\beta_i\alpha_{i}^{-1}| = |\gamma_{n,n}^{-1}\beta_i|$ implies that $\beta_i\leq \alpha_i\leq\gamma_{n,n}$. Thus, we have established Condition \eqref{Fnn-min-3}.
\end{proof}

The main result of this section is the following theorem:

\begin{thm}\label{master-ineq}
For any sequence of permutations $\{ \beta_i \}_{i=1}^k$ in $\mathcal{S}_{2n},$ if at least one of them belongs to $\mathcal{S}_{2n,c},$ then we have 
\begin{equation*}
\bfX (2n-2)-F_{n,n}(\beta)\leq -2.
\end{equation*}
	\end{thm}

\begin{proof}
Note that the maximum flow problem can be solved by augmenting paths, we consider the following two cases.
			
			\textbf{Case 1:} If there exists $i_0\in[k]$, such that the flow $\bff = \{(\bfp_i,\bfX_i)\}_{i=1}^l$ with $|\bff| = \bfX$ includes a augmenting path $\bfp_j$ with residual capacity $\bfX_j$ as follows:
			\begin{equation*}
				\bfp_j:\ \gamma_{n,n}\to\beta_{j_1}\to \cdots\to\beta_{j_{l^\prime(j)}}\to\beta_{i_0}\to\beta_{j_{l^\prime(j)+1}}\to\cdots\to\beta_{j_{l(j)}}\to\Id_{2n},
			\end{equation*}
and $\beta_{i_0} \in \mathcal{S}_{2n, c}.$ After summing up all the inequalities coming from other augmenting paths (except the path $\bfp_j$) in this solution, one has:
			\begin{equation*}
				\begin{split}
				F_{(n,n)}(\beta) &\geq (\bfX -\bfX_j)(2n-2)\\
				&\quad +\bfX_j\left[|\gamma_{n,n}^{-1}\beta_{j_1}|+|\beta_{j_1}^{-1}\beta_{j_2}|+\cdots+|\beta_{j_{l^\prime(j)}}^{-1}\beta_{i_0}|+|\beta_{i_0}^{-1}\beta_{j_{l^\prime(j)+1}}|\cdots+|\beta_{j_{l(j)}}|\right]\\
				&\geq (\bfX -\bfX_j)(2n-2)+\bfX_j(|\gamma_{n,n}^{-1}\beta_{i_0}|+|\beta_{i_0}|).
				\end{split}
			\end{equation*}
			
Note that $\beta_{i_0}$ can be decomposed into the product of transpositions $\tau_1, \ldots, \tau_p$, i.e., $\beta_{i_0} = \tau_1 \cdots \tau_{p}$. Then, one of these transpositions, $\tau_{p_0}$, connects $\{1, \ldots, n\}$ and $\{n+1, \ldots, 2n\}$. Using \cite[Lemma 23.10]{NS2006}, we obtain
			\begin{equation*}
				\#(\gamma_{n,n}^{-1}\tau_1\cdots\tau_{p_0})\leq \#(\gamma_{n,n}^{-1})-1+(p_0-1).
			\end{equation*}
It follows that
\begin{equation*}
\#(\gamma_{n,n}^{-1}\tau_1\cdots\tau_p)\leq \#(\gamma_{n,n}^{-1})-1+(p-1),
\end{equation*}
which is equivalent to 
\begin{equation*}
\#(\gamma_{n,n}^{-1}\tau_1\cdots\tau_p)-2n\leq \#(\gamma_{n,n}^{-1})-2n-1+(|\beta_{i_0}|-1),
\end{equation*}
and thus, 
\begin{equation*}
2n\leq |\gamma_{n,n}^{-1}\beta_{i_0}|+|\beta_{i_0}|.
\end{equation*}
Therefore, 
\begin{equation*}
F_{(n,n)}(\beta)\geq (\bfX-\bfX_{j})(2n-2)+\bfX_{j}\cdot 2n\geq\bfX (2n-2)+2.
\end{equation*}
			
\textbf{Case 2:} If for any flow $\bff = \{(\bfp_i,\bfX_i)\}_{i=1}^l$ with $|\bff| = \bfX$, all the augmenting path $\{\bfp_i\}_{i=1}^l$ do not include any connected permutations. One can only find connected permutation in the residual network. Let's start from inequality \eqref{ine-res}:
			\begin{equation*}
				\begin{split}
				F_{n,n}(\beta)\geq& \bfX (2n-2) + \sum_{(\beta_i,\Id_{2n})\in\bfE_{\bff}}\bfc_\bff(\beta_{i},\Id_{2n})|\beta_i|\\
				&\ +\sum_{(\gamma_{n,n},\beta_i)\in\bfE_{\bff}}\bfc_\bff(\gamma_{n,n},\beta_i)|\gamma_{n,n}^{-1}\beta_i|+\sum_{(\beta_i,\beta_j)\in\bfE_{\bff},i<j}\bfc_\bff(\beta_i,\beta_j)|\beta_i^{-1}\beta_j|.
				\end{split}
			\end{equation*}
			
We assume that $\beta_{i_0} \in \mathcal{S}_{2n , c}.$ Since the vertex $\beta_{i_0}$ does not appear in any augmenting path, one has 
\begin{equation*}
\bfc_\bff(\beta_{i_0},\Id_{2n}) = \bfc(\beta_{i_0},\Id_{2n}), \; \bfc_\bff(\gamma_{n,n},\beta_{i_0}) = \bfc(\gamma_{n,n},\beta_{i_0}), \; \text{and} \; \bfc_\bff(\beta_{i_0},\beta_j) = \bfc(\beta_{i_0},\beta_{j})
\end{equation*}
 for all $1 \leq j \leq k,$ $j\not=i_0$. Because
 \begin{equation*}
 \sum_{\substack{1\leq j\leq k\\j\not=i_0}}\bfc_\bff(\beta_{i_0},\beta_{j}) = \sum_{\substack{1\leq j\leq k\\j\not=i_0}}\bfc(\beta_{i_0},\beta_{j}) = \sum_{\substack{1\leq j\leq k\\j\not=i_0}}e_{i_0,j} > 0,
 \end{equation*} 
 there exists $i_1\not= i_0$ such that $\bfc_\bff(\beta_{i_0},\beta_{i_1}) > 0.$ If $\bfc_{\bff}(\beta_{i_0},\Id_{2n}) > 0$ and $\bfc_{\bff}(\gamma_{n,n},\beta_{i_0}) > 0$, then there exists one more augmenting path $\gamma_{n,n}\to\beta_{i_0}\to\Id_{2n}$ in $\bfG_\bff$, which is a contradiction. On the other hand, the assumptions $\bfc(\beta_i,\Id_{2n})+\bfc(\gamma_{n,n},\beta_i) = s_i+t_i > 0$ for all $1\leq i\leq k$, 
 imply that 
 $\bfc_\bff(\beta_{i_0},\Id_{2n}) = \bfc_{\bff}(\gamma_{n,n},\beta_{i_0}) = 0$ is also impossible.
In summary, we claim that one of $\bfc_{\bff}(\beta_{i_0},\Id_{2n})$ and $\bfc_{\bff}(\gamma_{n,n},\beta_{i_0})$ is zero, while another one is greater than zero.

Now suppose that $\bfc_{\bff}(\beta_{i_0},\Id_{2n})>0$ and $\bfc_{\bff}(\gamma_{n,n},\beta_{i_0})=0.$ 
If $|\beta_{i_0}^{-1}\beta_{i_1}| \geq 1$, 
then we have
\begin{equation*}
F_{n,n}(\beta) \geq  \bfX (2n-2) + \bfc_\bff(\beta_{i_0},\beta_{i_1})|\beta_{i_0}^{-1}\beta_{i_1}| + \bfc_{\bff}(\beta_{i_0},\Id_{2n})|\beta_{i_0}| \geq \bfX (2n-2) + 2
\end{equation*}
If $|\beta_{i_0}^{-1}\beta_{i_1}|=0$, i.e., $\beta_{i_1} = \beta_{i_0} \in \mathcal{S}_{2n ,c}.$ Then we have 
\begin{equation*}
F_{n,n}(\beta) \geq \bfX (2n-2) + \bfc_{\bff}(\beta_{i_0},\Id_{2n})|\beta_{i_0}| + \bfc_{\bff}(\beta_{i_1},\Id_{2n})|\beta_{i_1}| \geq \bfX (2n-2) + 2.
\end{equation*}
Similarly, for the case $\bfc_{\bff}(\beta_{i_0},\Id_{2n})=0$ and $\bfc_{\bff}(\gamma_{n,n},\beta_{i_0})>0,$ we can obtain the same bound. 
Thus the theorem is proved.
\end{proof}

\begin{cor}\label{master-ineq-general}
For any $\{\alpha_i\}_{i=1}^k$ and $\{ \beta_i \}_{i=1}^k$ in $\mathcal{S}_{2n},$ if at least one of $\alpha_1, \alpha_2, \ldots, \alpha_k, \beta_1, \beta_2, \ldots, \beta_k$ belongs to $\mathcal{S}_{2n, c}$, then we have 
\begin{equation*}
		\bfX (2n-2) - F_{n,n}(\alpha,\beta) \leq -2.
\end{equation*}
\end{cor}
	
\begin{proof}
If one of $\alpha_1, \alpha_2, \ldots, \alpha_k$ is connected, by the inequality $|\beta_i \alpha_i^{-1}| + |\beta_i\beta_j^{-1}| + |\beta_j\alpha_j^{-1}| \geq |\alpha_i\alpha_j^{-1}|$, we have
		\begin{equation*}
			\begin{split}
				F_{n,n}(\alpha,\beta) = & \sum_{i=1}^{k} \left(s_i|\gamma_{n,n}^{-1}\alpha_i|+t_i|\alpha_i|\right) + \sum_{1\leq i < j\leq k}e_{i,j}|\beta_i^{-1}\beta_j| + \sum_{i=1}^{k}d_i|\beta_i\alpha_i^{-1}|\\
				\geq & \sum_{i=1}^{k} \left(s_i|\gamma_{n,n}^{-1}\alpha_i|+t_i|\alpha_i|\right) + \sum_{1\leq i < j\leq k}e_{i,j}|\alpha_i\alpha_j^{-1}|\\
				=& F_{n,n}(\alpha).
			\end{split}
		\end{equation*}
Thus, by Theorem \ref{master-ineq}, we obtain our result.
	
On the other hand, suppose that one of $\beta_1, \beta_2, \ldots, \beta_k$ is connected. Combining \eqref{eq:inequality} and Theorem \ref{master-ineq}, we can conclude our proof.
	\end{proof}

	
	
\section{Proof of Theorem \ref{thm:almost-sure}}
Recall that the network $\tilde{\bf G}$ and ${\bf G}$ are given in Subsection \ref{subsec:flownetwork} and \ref{subsec:min}, respectively. Let $\tilde{\bfX}$ be the maximal value of the flow in the network $\tilde{\bfG}$ that can be sent from the source $\gamma_{n}$ to the sink $\Id_{n}$, and let ${\bfX}$ be the maximal value of the flow in the network $\bfG$ that can be sent from the source $\gamma_{n, n}$ to the sink $\Id_{2n}.$ It is clear that $\tilde{\bfG}$ is isomorphic to $\bfG$. Let $\mathbf{F}$ be the map from $\bfG$ to $\tilde{\bfG}$ defined as $\mathbf{F}(\Id_{2n}) = \Id_n,$ $\mathbf{F}(\gamma_{n,n}) = \gamma_n$, and $\mathbf{F}(\beta_i)= \tilde{\beta}_i,$ for all $1\leq i\leq k$.

For any maximal flow $\bff_1 = \{(\bfp_i,\bfX_i)\}_{i=1}^l$ in $\bfG$, with the augmenting paths 
\begin{equation*}
\bfp_i:\gamma_{n,n}\to \beta_{i_{1}}\to\beta_{i_{2}}\to\cdots\to\beta_{i_{l(i)}}\to\Id_{2n}, \ 1\leq i\leq l,
\end{equation*}
there exists a flow $\tilde{\bff}_1 = \{(\tilde{\bfp}_i,\bfX_i)\}_{i=1}^l$ in $\tilde{\bf G},$ such that
\begin{equation*}
\tilde{\bfp}_i:\mathbf{F}(\gamma_{n,n})\to \mathbf{F}(\beta_{i_{1}})\to\mathbf{F}(\beta_{i_{2}})\to\cdots\to\mathbf{F}(\beta_{i_{l(i)}})\to\mathbf{F}(\Id_{2n}) 
\end{equation*}
for all $1\leq i\leq l$. Then we have $\tilde{\bfX}\geq\bfX$. On the other hand, for any maximal flow $\tilde{\bff}_2 = \{(\tilde{\bfp}_j,\tilde{\bfX}_j)\}_{j=1}^{l^\prime}$ in $\tilde{\bfG},$
with the augmenting paths
\begin{equation*}
\tilde{\bfp}_j:\gamma_{n}\to \bet_{j_{1}}\to\bet_{j_{2}}\to\cdots\to\bet_{j_{l^\prime(j)}}\to\Id_{n},\ 1\leq j\leq l^\prime,
\end{equation*}
there exist one flow $\bff_2 = \{(\bfp_j,\tilde{\bfX}_j)\}_{j=1}^{l^\prime}$ in $\bfG$, such that
\begin{equation*}
\bfp_j:\mathbf{F}^{-1}(\gamma_{n})\to \mathbf{F}^{-1}(\bet_{j_{1}})\to\mathbf{F}^{-1}(\bet_{j_{2}})\to\cdots\to\mathbf{F}^{-1}(\bet_{j_{l^\prime(j)}})\to\mathbf{F}^{-1}(\Id_{n})
\end{equation*}
for all $1\leq j\leq l^\prime$. We get $\bfX\geq\tilde{\bfX}$. Combining the Theorem \ref{flow-area}, it is convinced that
\begin{equation}\label{flow-equal}
\bfX= \tilde{\bfX} = |\partial \mathcal{S}|.
\end{equation}

\begin{rem}
For the fattened graph $\mathcal{G},$ one can also obtain a flow network, still denote by $\mathcal{G}$. The vertices are still the vertices of $\mathcal{G}$, and the capacities $\bfc$ are given by 
\begin{equation*}
\bfc(\mathcal{V}_i,\mathcal{V}_j) = \begin{cases}
			e_{i,j} & \text{ if } i\not= j,\\
			0 & \text{ if } i = j.
		\end{cases} 
\end{equation*}
Add two distinguished vertices, called $\Id$ and $\gamma$. The remaining capacities are given by $\bfc(\mathcal{V}_i,\Id) = t_i$ and $\bfc(\gamma,\mathcal{V}_i) = s_i$. It is obvious that $\bfG$, $\tilde{\bfG}$ and $\mathcal{G}$ are three isomorphic flow networks. Surely, they have the same maximum value of flow.			
\end{rem}

\begin{lem}\label{lem:disconnect}
Suppose that $\alpha_1,\ldots,\alpha_k, \beta_1,\ldots,\beta_k \notin S_{2n, c},$ denote
\begin{equation*}
\alpha_i : = (\alpha_i^{(1)}, \alpha_i^{(2)}) \; \text{and} \; \beta_i : = (\beta_i^{(1)}, \beta_i^{(2)}), i=1, \ldots, k,
\end{equation*}
where $\alpha_i^{(s)}, \beta_i^{(s)} \in S_n, s=1,2.$  
Then, we have 
\begin{equation*}
F_{n, n} (\alpha, \beta) = F_n (\alpha^{(1)}, \beta^{(1)})+ F_n (\alpha^{(2)}, \beta^{(2)}).
\end{equation*}
\end{lem}

\begin{proof}
By a directly computation, we can prove this lemma. We omit the details. 
\end{proof}

Now we are ready to prove Proposition \ref{prop:almost surely}.
\begin{proof}[Proof of Proposition \ref{prop:almost surely}]
Note that $\bfX= \tilde{\bfX}.$ It follows from \eqref{eq:uni-cov-1.1} and \eqref{eq:uni-cov-2} that 
\begin{equation}\label{eq:1}
				\begin{split}
					 \mathbb{E}\bigg[\bigg(\frac{1}{N^{\tilde{\bfX}} }\Tr\big[N^{\tilde{\bfX}} \rho_\mathcal{S}\big]^n\bigg)^2\bigg]
					  = \sum_{\substack{\alpha_1,\ldots,\alpha_k,\\\beta_1,\ldots,\beta_k\in\mathcal{S}_{2n}}}N^{\tilde{\bfX} (2n-2)-F_{n,n}(\alpha,\beta)}\bigg(\prod_{i=1}^{k}\phi(\beta_i\alpha_i^{-1})+O(N^{-2})\bigg),
				\end{split}
			\end{equation}
and
\begin{equation}\label{eq:2}
			\begin{split}
 \bigg[\frac{1}{N^{\tilde{\bfX}} }\mathbb{E}\bigg(\Tr\big[N^{\tilde{\bfX}} \rho_\mathcal{S}\big]^n\bigg)\bigg]^2
				&= \sum_{\substack{\alpha_1^{(1)},\ldots,\alpha_k^{(1)},\beta_1^{(1)},\ldots,\beta_k^{(1)},\\ \alpha_1^{(2)},\ldots,\alpha_k^{(2)}, \beta_1^{(2)},\ldots, \beta_k^{(2)}\in\mathcal{S}_n}}N^{\tilde{\bfX} (2n-2)-F_n(\alpha^{(1)},\beta^{(1)})-F_n(\alpha^{(2)},\beta^{(2)})}\\
				& \quad \times \bigg(\prod_{i=1}^{k}\phi(\beta_i^{(1)}(\alpha_i^{(1)})^{-1})\phi(\beta_i^{(2)}(\alpha_i^{(2)})^{-1})+O(N^{-2})\bigg).
			\end{split}
			\end{equation}
By Proposition \ref{Fn min cond} and Proposition \ref{Fnn min cond}, we have 
\begin{equation*}
\bfX (2n-2)-F_{n,n}(\alpha,\beta) \leq 0,
\end{equation*}
and 
\begin{equation*}
\tilde{\bfX} (2n-2)-F_n(\alpha^{(1)},\beta^{(1)})-F_n(\alpha^{(2)},\beta^{(2)}) \leq 0.
\end{equation*}
Moreover, for any $\alpha = (\alpha^{(1)}, \alpha^{(2)}), \beta = (\beta^{(1)}, \beta^{(2)}) \notin \mathcal{S}_{2n, c},$ we have 
\begin{equation*}
\phi(\beta \alpha^{-1}) = \phi (\beta^{(1)} (\alpha^{(1)})^{-1}) \cdot \phi (\beta^{(2)} (\alpha^{(1)})^{-2}).
\end{equation*}
Hence, combining Lemma \ref{lem:disconnect} and Corollary \ref{master-ineq-general}, we can deduce that 
\begin{equation*}
	\begin{split}
\mathbb{E}\bigg[\bigg(\frac{1}{N^{\tilde{\bfX}} }\Tr\big[N^{\tilde{\bfX}} \rho_\mathcal{S}\big]^n\bigg)^2\bigg] & -\bigg[\frac{1}{N^{\tilde{\bfX}} }\mathbb{E}\bigg(\Tr\big[N^{\tilde{\bfX}} \rho_\mathcal{S}\big]^n\bigg)\bigg]^2\label{q2-1}\\
		&=\sum_{\substack{\alpha_1,\cdots,\alpha_k,\beta_1,\cdots,\beta_k\in\mathcal{S}_{2n},\\\text{at least one of them is in }\mathcal{S}_{2n,c}}}N^{\bfX (2n-2)-F{n,n}(\alpha,\beta)}\bigg(\prod_{i=1}^{k}\phi(\beta_i\alpha_i^{-1})+O(N^{-2})\bigg)\\
& = O(N^{-2}).
\end{split}
\end{equation*}
The rest of the proof is standard, by the Chebyshev inequality and the Borel–Cantelli lemma, we can complete our proof.
\end{proof}

\begin{lem}[Proposition 4.4 \cite{BI12}]\label{lem:convergence}
Let $f$ be a continuous function on $\mathbb{R}$ with polynomial growth and $\nu_N$ be a sequence of probability measures which converges in moments to a compactly supported measure $\nu.$ Then, $\int_{\mathbb{R}} f d \nu_N \rightarrow \int_{\mathbb{R}} f d \nu$ as $N \rightarrow \infty.$ 
\end{lem}
We end this section by proving Theorem \ref{thm:almost-sure}.

\begin{proof}[Proof of Theorem \ref{thm:almost-sure}]
By Proposition \ref{prop:almost surely}, almost surely as $N \rightarrow \infty$, $\mu_{N}$ converges in moments to $\mu$. Then, applying Lemma \ref{lem:convergence} for $f(t) = t \log t$, almost surely as $N\to\infty$,
\begin{equation*}
\begin{split}
\int_{\mathbb{R}} t\log td\mu + o(1) &= \int_{\mathbb{R}} x \log x d \mu_N = \frac{1}{N^{\tilde{\bfX}} }\Tr\left(N^{\tilde{\bfX}} \rho_\mathcal{S}\log\left(N^{\tilde{\bfX}} \rho_\mathcal{S}\right)\right) \\
&= \Tr(\rho_\mathcal{S}\log\rho_\mathcal{S}) + \tilde{\bfX} \log N \\
& = - H(\rho_{\mathcal{S}}) +  |\partial\mathcal{S}| \log N.
\end{split}
\end{equation*}
\end{proof}
		
	

\section{Examples}\label{sec:example}
In this section, we apply our results to some concrete examples, demonstrating the application of random graph states and their connections with the free Poisson distribution.

	First, we revisit some known results about the free Poisson (or Marchenko–Pastur) distribution. Let $A_{M,N}$ be an rectangular $M\times N$ random matrix with independent Gaussian entries, where the entries $a_{ij}$ are standard complex Gaussian random variables with mean $0$ and $\mathbb{E}(|a_{ij}|^2)=1$, and define $X_N = \frac{1}{N}A_{M,N}^\ast A_{M,N}$. Assuming that $M$ and $N$ tend to infinity in such a way that the ratio $M/N$ converges to a finite limit $c>0$. Then, the empirical eigenvalue distribution of $X_N$ converges weakly to a Free Poisson distribution with rate $c$ is given by 
 $$d\pi_{c}(t) =\max\{1-c,0\}\delta_0+\frac{\sqrt{4c-(t-1-c)^2}}{2\pi t}\mathbf{1}_{[1+c-2\sqrt{c},1+c+2\sqrt{c}]}(t)dt,$$
	where $\mathbf{1}_A$ denotes the indicator function of the set $A$. It is known that the mean entropy of this probability distribution is given by:
	\begin{equation}
		H(\pi_{c}) = -\int t\log td\pi_{c}(t) =
		\begin{cases}
			-\frac{1}{2}-c\log (c) & \text{if } c \geq 1,\\
			-\frac{c^2}{2} & \text{if } 0<c<1.
		\end{cases}
	\end{equation}
	For further details on this distribution, we refer to \cite{AD09,TS11,BIK0}.

	
	\subsection{One dimensional chain}
The chain graph (see Figure \ref{Chain Case-1}) has $4$ vertices with degrees $d_1 = d_4 = 1$ and $d_2 = d_3 = 2$. The edges in this graph satisfy $e_{1,2} = e_{2,3} = e_{3,4} = 1$. The corresponding fattened graph is shown in Figure \ref{Chain Case-2}, where $\mathcal{S}$ is the shadow region, and the other region is $\mathcal{T}$. The marginal is obtained by partially tracing the vertices marked with "$\times$". Hence $|\partial\mathcal{S}| = 3$.
\begin{figure}[htbp]
\begin{minipage}[b]{.5\textwidth}
    \begin{tikzpicture}
			[roundnode/.style={circle, draw=black!100, fill=green!20, thick, minimum size=1mm},]
			\node[roundnode] (c1) at (-1.5,0) {};
			\node[roundnode] (c2) at (-0.5,0) {};
			\node[roundnode] (c3) at (0.5,0) {};
			\node[roundnode] (c4) at (1.5,0) {};
			\node (P1) at (-1.5,-0.5) {$v_{1}$};
			\node (P2) at (-0.5,-0.5) {$v_{2}$};
			\node (P3) at (0.5,-0.5) {$v_{3}$};
			\node (P4) at (1.5,-0.5) {$v_{4}$};
			\draw (c1)--(c2)--(c3)--(c4);
		\end{tikzpicture}
		\caption{One dimensional chain.} \label{Chain Case-1}
\end{minipage} %
\begin{minipage}[b]{.5\textwidth}
  \centering
		\begin{tikzpicture}
			[roundnode/.style={circle, draw=black!100, fill=green!20, thick, minimum size=1mm},]
			\filldraw[fill=yellow!20!,even odd rule]
			(-2,-0.6) rectangle (1,0.6);
			\node[roundnode] (c1) at (-2.75,0) {};
			\node[roundnode] (c2) at (-1.25,0) {};
			\node[roundnode] (c3) at (-0.75,0) {};
			\node[roundnode] (c4) at (0.75,0) {};
			\node[roundnode] (c5) at (1.25,0) {};
			\node[roundnode] (c6) at (2.75,0) {};
                \node (T1) at (-2.75,0) {$\times$};	
                \node (T2) at (0.75,0) {$\times$};	
                \node (T3) at (2.75,0) {$\times$};	
                \node (S) at (0,-0.3) {$\mathcal{S}$};
                \node (T) at (2,0.3) {$\mathcal{T}$};
			\draw (-2.75,0) circle (0.5);
			\draw (-1,0) ellipse (0.75 and 0.5);
			\draw (1,0) ellipse (0.75 and 0.5);
			\draw (2.75,0) circle (0.5);
			\node (P1) at (-2.75,-0.9) {$\mathcal{V}_1$};
			\node (P2) at (-1,-0.9) {$\mathcal{V}_{2}$};
			\node (P3) at (1,-0.9) {$\mathcal{V}_{3}$};
			\node (P4) at (2.75,-0.9) {$\mathcal{V}_{4}$};

			\draw (c2)--(c1);
			\draw (c4)--(c3);
			\draw (c6)--(c5);
		\end{tikzpicture}
		\caption{Chain Case (Fattened Graph).} \label{Chain Case-2}
\end{minipage}%
\begin{minipage}[b]{.5\textwidth}
  \centering
		\begin{tikzpicture}
			[roundnode/.style={circle, draw=black!100, fill=green!0, thick, minimum size=5mm},]
			\node[roundnode] (c0) at (-3,0.9) {$\gamma_{n,n}$};
			\node[roundnode] (c1) at (-3,-0.9) {$\beta_1$};
			\node[roundnode] (c2) at (-1,-0.3) {$\beta_2$};
			\node[roundnode] (c3) at (1,0.3) {$\beta_3$};
			\node[roundnode] (c4) at (3,0.9) {$\beta_4$};
			\node[roundnode] (cc) at (3,-0.9) {$\Id_{2n}$};
			\node (s01) at (-1.7,0.3) {$2$};
			\node (s02) at (-1,0.8) {$1$};
			\node (s1) at (-2,-0.4) {$1$};
			\node (s2) at (0,-0.2) {$1$};
			\node (s3) at (2,0.8) {$1$};
			\node (s11) at (0,-1.1) {$1$};
			\node (s12) at (2,-0.1) {$1$};
			\node (s13) at (3.1,0) {$1$};
			\draw (c0)--(c2)--(c1);
			\draw (c0)--(c3)--(c2);
			\draw (cc)--(c1);
			\draw (cc)--(c3)--(c4);
			\draw (cc)--(c4);
		\end{tikzpicture}
		\caption{Chain Case (Flow Network).} \label{Chain Case-3}
\end{minipage}
\end{figure}
Then we can write the function $F_{n,n}(\beta)$ as follows:
    \begin{equation}\label{F-nn chain}
	\begin{split}
F_{n,n}(\beta) = & \sum_{i=1}^{k} \left(s_i|\gamma_{n,n}^{-1}\beta_i|+t_i|\beta_i|\right) + \sum_{1\leq i < j\leq k}e_{i,j}|\beta_i^{-1}\beta_j|\\
           = & |\beta_1| + 2|\gamma_{n,n}^{-1}\beta_2| + |\beta_3| + |\gamma_{n,n}^{-1}\beta_3| + |\beta_4| + |\beta_1^{-1}\beta_2| +|\beta_2^{-1}\beta_3| + |\beta_3^{-1}\beta_4|.
           \end{split}
		\end{equation}

 Figure \ref{Chain Case-3} is the flow network constructed from Equation \eqref{F-nn chain}. For every edge, the nearby number is the capacities of that edge. 
 There are two maximal flow $\bff_1$ and $\bff_2$ in this flow network:
 \begin{equation}
 \begin{split}
     \bff_1 & = \begin{cases}
         \gamma_{n,n}\to\beta_2\to\beta_1\to\Id_{2n}\text{ with residual capacity }1,\\
         \gamma_{n,n}\to\beta_2\to\beta_3\to\Id_{2n}\text{ with residual capacity }1,\\
         \gamma_{n,n}\to\beta_3\to\beta_4\to\Id_{2n}\text{ with residual capacity }1,\\
     \end{cases}
     \\
     \bff_2 & = 
     \begin{cases}
         \gamma_{n,n}\to\beta_2\to\beta_1\to\Id_{2n}\text{ with residual capacity }1,\\
         \gamma_{n,n}\to\beta_2\to\beta_3\to\beta_4\to\Id_{2n}\text{ with residual capacity }1,\\
         \gamma_{n,n}\to\beta_3\to\Id_{2n}\text{ with residual capacity }1.\\
     \end{cases}
     \end{split}
 \end{equation}
    Then we get the maximum value of flow in this network (Figure \ref{Chain Case-3}) is $\bfX = 3$. 
 $F_{n,n}(\beta)$ achieves its minimal value if and only if 
 \begin{equation}
     \begin{cases}
         \Id_{2n}\leq \beta_1\leq \beta_2\leq \gamma_{n,n},\\
         \Id_{2n}\leq \beta_4\leq \beta_3\leq \beta_2\leq  \gamma_{n,n}.
     \end{cases}
 \end{equation}
 On the other hand, suppose that $\beta_1$ is connected. By considering the augmenting paths of maximal flow $\bff_1$, we have
 \begin{equation}
     \begin{split}
         F_{n,n}(\beta) = & (|\gamma_{n,n}^{-1}\beta_2| + |\beta_1^{-1}\beta_2| + |\beta_1|) + (|\gamma_{n,n}^{-1}\beta_2| +|\beta_2^{-1}\beta_3|  + |\beta_3|) + (|\gamma_{n,n}^{-1}\beta_3| + |\beta_4| +   + |\beta_3^{-1}\beta_4|)\\
         \geq & |\gamma_{n,n}^{-1}\beta_1| + |\beta_1| + 2(2n-2)\\
         \geq & 3(2n-2) + 2.
     \end{split}
 \end{equation}
 
    By using \cite[Theorem 5.4]{BIK0}, the limit moments of the empirical eigenvalue measure of $N^3\rho_\mathcal{S}$ are:
	\begin{equation}
		\begin{split}
		\lim_{N\to\infty}\mathbb{E}\frac{1}{N^3}\Tr[(N^3\rho_\mathcal{S})^n] = \sum_{\Id_n\leq \tilde{\beta}_3\leq \gamma_n}1 = \int t^n d\pi_{1}(t).
		\end{split}
	\end{equation}
	Here $\pi_{1}$ is the Free Poisson probability measure with rate $1$. Combining the Proposition \ref{prop:almost surely} and Theorem \ref{thm:almost-sure}, one has $$H(\rho_\mathcal{S}) = 3\log N - \frac{1}{2} + o(1),$$
	almost surely as $N\to\infty$.

	\subsection{Two-dimensional lattice}
	Let us consider the two-dimensional lattice case, as shown in Figure \ref{Lattice Case}. $\mathcal{S}$ is the shadow region, the other region is $\mathcal{T}$. The marginal is obtained by partial tracing the vertices marked with "$\times$". Thus, we have $|\partial\mathcal{S}| = 6$. 

	\begin{figure}[htbp]
		\centering
  \begin{minipage}[b]{.5\textwidth}
		\begin{tikzpicture}
			[roundnode/.style={circle, draw=black!100, fill=green!20, thick, minimum size=1mm},]
   
                \node[roundnode] (c1) at (0,0,0) {};
                \node[roundnode] (c2) at (0,1,0) {};
                \node[roundnode] (c3) at (0,2,0) {};
                \node[roundnode] (c4) at (1,0,0) {};
                \node[roundnode] (c5) at (1,1,0) {};
                \node[roundnode] (c6) at (1,2,0) {};
                \node[roundnode] (c7) at (2,0,0) {};
                \node[roundnode] (c8) at (2,1,0) {};
                \node[roundnode] (c9) at (2,2,0) {};
			\node (P1) at (0,0,-0.5) {$v_7$};
			\node (P2) at (0,1,-0.5) {$v_4$};
			\node (P3) at (0,2,-0.5) {$v_1$};
			\node (P4) at (1,0,-0.5) {$v_8$};
                   \node (P5) at (1,1,-0.5) {$v_5$};
                   \node (P6) at (1,2,-0.5) {$v_2$};
                   \node (P7) at (2,0,-0.5) {$v_9$};
                   \node (P8) at (2,1,-0.5) {$v_6$};
                   \node (P9) at (2,2,-0.5) {$v_3$};
                

            \draw (c1)--(c2)--(c3)--(c6)--(c5)--(c4)--(c7)--(c8)--(c9)--(c6)--(c3);
			\draw (c1)--(c4)--(c7);
			\draw (c2)--(c5)--(c8);
		\end{tikzpicture}
		\caption{Two-dimensional lattice.} \label{Lattice Case}
\end{minipage} %
\begin{minipage}[b]{.4\textwidth}
  \centering
		\begin{tikzpicture}
			[roundnode/.style={circle, draw=black!100, fill=green!20, thick, minimum size=1mm}, roundnode-1/.style={circle, draw=black!100,  thick, minimum size=11mm}]
		   	\filldraw[fill=yellow!20!,even odd rule]
			(-2,0) -- (-1,1) -- (1,1) -- (2,0) -- (1,-1) -- (-1,-1) -- cycle;
			\node[roundnode-1] (C1) at (-2,2) {};
			\node[roundnode-1] (C2) at (0,2) {};
			\node[roundnode-1] (C3) at (2,2) {};
			\node[roundnode-1] (C4) at (-2,0) {};
			\node[roundnode-1] (C5) at (0,0) {};
			\node[roundnode-1] (C6) at (2,0) {};
			\node[roundnode-1] (C7) at (-2,-2) {};
			\node[roundnode-1] (C8) at (0,-2) {};
			\node[roundnode-1] (C9) at (2,-2) {};

			\node[roundnode] (c11) at (-1.7,2) {};
			\node[roundnode] (c10) at (-2,1.7) {};
			\node[roundnode] (c21) at (-0.3,2) {};
			\node[roundnode] (c22) at (0,1.7) {};
			\node[roundnode] (c23) at (0.3,2) {};
			\node[roundnode] (c31) at (1.7,2) {};
			\node[roundnode] (c32) at (2,1.7) {};
			\node[roundnode] (c41) at (-2,0.3) {};
			\node[roundnode] (c42) at (-1.7,0) {};
			\node[roundnode] (c43) at (-2,-0.3) {};
                \node (T5) at (-2,-0.3) {$\times$};
			\node[roundnode] (c51) at (-0.3,0) {};
			\node[roundnode] (c52) at (0,0.3) {};
			\node[roundnode] (c53) at (0.3,0) {};
			\node[roundnode] (c54) at (0,-0.3) {};
			\node[roundnode] (c61) at (1.7,0) {};
			\node[roundnode] (c62) at (2,0.3) {};
			\node[roundnode] (c63) at (2,-0.3) {};
		      \node (T6) at (2,0.3) {$\times$};
			\node[roundnode] (c71) at (-2,-1.7) {};
			\node[roundnode] (c72) at (-1.7,-2) {};
			\node[roundnode] (c81) at (-0.3,-2) {};
			\node[roundnode] (c82) at (0,-1.7) {};
			\node[roundnode] (c83) at (0.3,-2) {};
			\node[roundnode] (c91) at (1.7,-2) {};
			\node[roundnode] (c92) at (2,-1.7) {};
			\node (P1) at (-1.4,1.5) {$\mathcal{V}_{1}$};
			\node (P2) at (0.6,1.5) {$\mathcal{V}_{2}$};
			\node (P3) at (2.6,1.5) {$\mathcal{V}_{3}$};
			\node (P4) at (-1.3,-0.4) {$\mathcal{V}_{4}$};
			\node (P5) at (0.6,-0.5) {$\mathcal{V}_{5}$};
			\node (P6) at (2.6,-0.5) {$\mathcal{V}_{6}$};
			\node (P7) at (-1.4,-2.5) {$\mathcal{V}_{7}$};
			\node (P8) at (0.6,-2.5) {$\mathcal{V}_{8}$};
			\node (P9) at (2.6,-2.5) {$\mathcal{V}_{9}$};
			\node (S) at (-0.7,0.7) {$\mathcal{S}$};
                \node (T) at (1.3,1.3) {$\mathcal{T}$};
                \node (T1) at (0,0.3) {$\times$};
                \node (T2) at (0,-0.3) {$\times$};
                \node (T3) at (-0.3,0) {$\times$};
                \node (T4) at (0.3,0) {$\times$};
			\draw (c11)--(c21);
			\draw (c23)--(c31);
			\draw (c10)--(c41);
			\draw (c22)--(c52);
			\draw (c32)--(c62);
			\draw (c42)--(c51);
			\draw (c53)--(c61);
			\draw (c43)--(c71);
			\draw (c54)--(c82);
			\draw (c63)--(c92);
			\draw (c72)--(c81);
			\draw (c83)--(c91);
		\end{tikzpicture}
		\caption{Lattice Case (Fattened graph).} \label{Lattice Case (Fattened graph)}
\end{minipage}%
\begin{minipage}[b]{.4\textwidth}
  \centering
\begin{tikzpicture}
			[roundnode/.style={circle, draw=black!100, fill=green!0, thick, minimum size=5mm},]
			\node[roundnode] (c00) at (-2.8,0) {$\beta_4$};
			\node[roundnode] (c01) at (0,0) {$\beta_5$};
			\node[roundnode] (c02) at (2.8,0) {$\beta_6$};
			\node[roundnode] (c10) at (-1.8,1) {$\beta_1$};
			\node[roundnode] (c11) at (1,1) {$\beta_2$};
			\node[roundnode] (c12) at (3.8,1) {$\beta_3$};
			\node[roundnode] (-c12) at (1.8,-1) {$\beta_9$};
			\node[roundnode] (-c11) at (-1,-1) {$\beta_8$};
			\node[roundnode] (-c10) at (-3.8,-1) {$\beta_7$};
			\node[roundnode] (c0) at (0,3) {$\Id_{2n}$};
			\node[roundnode] (c1) at (0,-2) {$\gamma_{n,n}$};	
			\color{blue}
			\draw (c0)..controls (-0.5,3) and (-1.8,2)..(c10);
			\draw (c0)..controls (0.3,2.9) and (1,2)..(c11);
			\draw (c0)..controls (1.5,3) and (3.5,2)..(c12);
			\draw (c0)..controls (-1,3) and (-2.8,2)..(c00);
			\draw (c0)..controls (1,2.9) and (2.8,1.7)..(c02);
			\draw (c0)..controls (-1.3,3) and (-3.8,2)..(-c10);
			\draw (c0)..controls (-0.5,2) and (-1,1)..(-c11);
			\draw (c0)..controls (1,2.8) and (1.8,2)..(-c12);
			\draw (c1)--(c01);
                \draw (c1)..controls (-1,-2) and (-2.8,-1.5)..(c00);
                \draw (c1)..controls (1,-2) and (2.8,-1.5)..(c02);
			\node (ss5) at (0.1,-1.2) {$4$};
                \node (ss4) at (-1.8,-1.8) {$1$};
                \node (ss6) at (1.8,-1.8) {$1$};
			\node (d1) at (-1.2,2) {$2$};
			\node (d2) at (0.6,2) {$3$};
			\node (d3) at (3.4,2) {$2$};
			\node (d4) at (-2.4,1.25) {$2$};
			\node (d6) at (2.7,1.25) {$2$};
			\node (d7) at (-3.4,0.5) {$2$};
			\node (d8) at (-0.75,0.5) {$3$};
			\node (d9) at (2,0.5) {$2$};
			\color{red}
			\draw (c00)--(c10)--(c11)--(c12)--(c02)--(c01)--(c00)--(-c10)--(-c11)--(-c12)--(c02);
			\draw (c11)--(c01)--(-c11);
			\node (s1) at (-0.4,0.8) {$1$};
			\node (s2) at (2.4,0.8) {$1$};
			\node (s3) at (-2.1,0.4) {$1$};
			\node (s4) at (0.7,0.4) {$1$};
			\node (s5) at (3.5,0.4) {$1$};
			\node (s6) at (-1.4,-0.2) {$1$};
			\node (s7) at (1.4,-0.2) {$1$};
			\node (s8) at (-3.1,-0.6) {$1$};
			\node (s9) at (-0.3,-0.6) {$1$};
			\node (s10) at (2.4,-0.6) {$1$};
			\node (s11) at (0.6,-1.2) {$1$};
			\node (s12) at (-2.2,-1.2) {$1$};
		\end{tikzpicture}\caption{Lattice Case (Flow Network).} \label{Lattice Case (Flow Network)}
\end{minipage}
\end{figure}

	Under our assumptions, the associated function (from Figure \ref{Lattice Case (Fattened graph)}) is given by
	\begin{equation}\label{lattice fnn}
		\begin{split}
		F_{n,n}(\beta)=&4|\gamma_{n,n}^{-1}\beta_5|+|\gamma_{n,n}^{-1}\beta_4|+|\gamma_{n,n}^{-1}\beta_6| +2(|\beta_1|+|\beta_3|+|\beta_4|+|\beta_6|+|\beta_7|+|\beta_9|)+3(|\beta_2|+|\beta_8|)\\
		&+|\beta_1\beta_2^{-1}|+|\beta_2\beta_3^{-1}|+|\beta_4\beta_5^{-1}|+|\beta_5\beta_6^{-1}|+|\beta_7\beta_8^{-1}|+|\beta_8\beta_9^{-1}|\\
		&+|\beta_1\beta_4^{-1}|+|\beta_4\beta_7^{-1}|+|\beta_2\beta_5^{-1}|+|\beta_5\beta_8^{-1}|+|\beta_3\beta_6^{-1}|+|\beta_6\beta_9^{-1}|.
		\end{split}
	\end{equation}
The maximum value of the flow in the flow network $\bfG$ (Figure \ref{Lattice Case (Flow Network)}) associated with $F_{n,n}(\beta)$ is $\bfX = 6$. It's worth noting that there exist multiple maximal flows in this network. We claim that: \textit{
	$F_{n,n}(\beta)$ gets the minimal value if and only if
\begin{equation}\label{lattice min}
\beta_1 = \beta_2 = \beta_3 = \beta_4 = \beta_6 = \beta_7 = \beta_8 = \beta_9 = \Id_{2n},\text{ and } \Id_{2n} \leq \beta_5 \leq \gamma_{n,n}.
\end{equation}
 }
\begin{proof}
    For any maximal flow $\bff$ in flow network $\bfG$ (in Figure \ref{Lattice Case (Flow Network)}), if the residual capacity $\{\bfc_{\bff}(\beta_i,\Id_{2n})\}_{1\leq i\leq 9, i\not=4,5,6}$ of $\bfG_{\bff}$ are all bigger than zero, by Proposition \ref{Fnn min cond}, $F_{n,n}(\beta)$ achieves its minimal value only if $$\beta_1=\beta_2=\beta_3=\beta_7=\beta_8=\beta_9=\Id_{2n}.$$

    If some of the residual capacity $\{\bfc_{\bff}(\beta_i,\Id_{2n})\}_{1\leq i\leq 9, i\not=4,5,6}$ of $\bfG_{\bff}$ are zeros, for example, suppose that $\bfc_{\bff}(\beta_1,\Id_{2n}) = 0$. There must exist one augmenting path
$$\bfp_1: \gamma_{n,n}\to\cdots\to\beta_2\to\beta_1\to\Id_{2n}$$
with residual capacity $1$ in $\bff$. By Proposition \ref{Fnn min cond}, we use the inequality
$$Id_{2n}\leq \beta_1\leq \beta_2\leq \cdots\leq \gamma_{n,n}$$
to get the $F_{n,n}(\beta)$ minimal value. Let $\bff_1$ be the flow updated from zero flow by $\bfp_1$. Since path $\bfp_1$ has pass the edge $(\beta_1,\beta_2)$, $\bfc_{\bff_1}(\beta_2,\Id_{2n}) = 3>2 \geq \sum_{\bfu\in \bfG, \bfu\not=\Id_{2n}}\bfc_{\bff_1}(\bfu,\beta_2).$ Hence 
$$\bfc_{\bff}(\beta_2,\Id_{2n})\geq\bfc_{\bff_1}(\beta_2,\Id_{2n})-\sum_{\bfu\in \bfG, \bfu\not=\Id_{2n}}\bfc_{\bff_1}(\bfu,\beta_2)>0,$$
i.e., $\beta_2 = \Id_{2n}$ when the $F_{n,n}(\beta)$ gets minimal value. We can say $\beta_1 = \beta_2 = \Id_{2n}$. Similarly, we have $F_{n,n}(\beta)$ gets minimal value only if $\beta_1 = \beta_2 = \beta_3 = \beta_7 = \beta_8 = \beta_9 = \Id_{2n}.$

Considering $\beta_4$, if $\bfc_{\bff}(\beta_4,\Id_{2n})>0$, then $\beta_4 = \Id_{2n}$ when $F_{n,n}(\beta)$ attains its minimal value. If $\bfc_{\bff}(\beta_4,\Id_{2n}) = 0$, given that $\bfc(\gamma_{n,n},\beta_4) = t_4 = 1<2 = s_4 = \bfc(\gamma_{n,n},\beta_4)$, there exists an augmenting path $\bfp_2$:
$$\bfp_2: \gamma_{n,n}\to\cdots\to\beta_i\to\beta_4\to\Id_{2n},\ i \in\{1,5,7\}.$$
If $i$ can be $1$ or $7$, we have $\beta_4\leq\beta_i = \Id_{2n}$. If $i$ can only be $5$, since $\bfc_{\bff}(\beta_4,\beta_7) > 0$ implies $\beta_4 = \beta_7$ directly when $F_{n,n}(\beta)$ attains the minimal value, let us consider $\bfc_{\bff}(\beta_4,\beta_7) = 0$. Then there must be an augmenting path $\bfp_3$,
$$\bfp_3: \gamma_{n,n}\to \cdots\to\beta_7\to\beta_4\to\beta_1\to\cdots\to\Id_{2n},$$
or
$$\bfp_3: \gamma_{n,n}\to \cdots\to\beta_1\to\beta_4\to\beta_7\to\cdots\to\Id_{2n}.$$
We can also say $\Id_{2n} = \beta_1\leq \beta_4\leq \beta_7 = \Id_{2n}$ or $\Id_{2n} = \beta_7\leq \beta_4\leq \beta_1 = \Id_{2n}.$ 

In summary, we can conclude $$\beta_1=\beta_2=\beta_3=\beta_4 = \beta_6 = \beta_7=\beta_8=\beta_9=\Id_{2n},\text{ and } \Id_{2n}\leq \beta_5 \leq \gamma_{n,n},$$
if $F_{n,n}(\beta)$ attains its minimal value.

Furthermore, by applying the above conditions and the inequality
\begin{equation}
|\gamma_{n,n}^{-1}\beta_{5}| + |\beta_{5}|\geq 2n-2,
\end{equation}
in $F_{n,n}(\beta)$, we can establish $F_{n,n}(\beta) = 6(2n-2)$, concluding our proof.
\end{proof}

Note that in Equation \eqref{lattice fnn}, for $1\leq i\leq 9,\ i\not= 5$, there is $2|\beta_i|$, and $2|\gamma_{n,n}^{-1}\beta_5|+2|\beta_5|$ is also included in this equation. If at least one of $\{\beta_i\}_{1\leq i\leq 9}$ is connected, then $F_{n,n}(\beta)\geq 6(2n-2) + 2$.

By employing \cite[Theorem 5.4]{BIK0}, the moments of the empirical eigenvalue measure of $N^6\rho_\mathcal{S}$ can be expressed as
\begin{equation}
\lim_{N\to\infty}\mathbb{E}\frac{1}{N^6}\Tr[(N^6\rho_\mathcal{S})^n]
= 1.
\end{equation}
Combining Proposition \ref{prop:almost surely} and Theorem \ref{thm:almost-sure}, we conclude that almost surely, as $N\to\infty$,
	\[ H\left(\rho_\mathcal{S}\right) = 6\log N+o(1).
	\]

\section{Conclusion}
In this work, we study the probability distribution of the reduced random graph state. Following the approach of Collins et. al. \cite{BIK3}, we have obtained a combinatoric inequality that involves connected permutations. As an application, we have estimated the variance of the von Neumann entropy of the reduced random graph state. Hence, we have proved the area law holds almost surely, which provides a positive answer to a question in \cite{BIK3}.

As mentioned in \cite[Remark 5.7]{BIK0}, the parameters of the graph state can be sampled by non-Hermitian complex Gaussian random ensembles. In fact, we can obtain similar results for this random model. A key point is that the Weingarten calculus is replaced instead by the Wick formula, and the same combinatoric inequality can be derived.



\vspace{2mm}
	
{\it Acknowledgments:} L. Zhao would like to thank Beno\^it Collins for his helpful discussion. We are partially supported by NSFC No. 12031004. 
	
	\bibliography{reference}

\end{document}